\documentclass[12pt]{amsart}
\usepackage{amssymb}

\usepackage{color}

\usepackage{latexsym}
\usepackage{epic, eepic, graphicx}
\usepackage{graphicx} \DeclareGraphicsExtensions{.ps}

\newcommand{\C}{{\mathbb C}}

\newcommand{\NN}{{\mathbb N}}

\newcommand{\RR}{{\mathbb R}}

\def\t2{{\mathbb T}^2}

\newcommand{\ZZ}{{\mathbb Z}}

\newcommand{\CI}{{\mathcal C}^\infty }
\newcommand{\CIc}{{\mathcal C}^\infty_{\rm{c}} }

\newcommand{\vol}{\operatorname{vol}}

\newcommand{\supp}{\operatorname{supp}}

\newcommand{\Spec}{\operatorname{Spec}}

\newcommand{\rest}{|}

\renewcommand{\Re}{\mathop{\rm Re}\nolimits}
\renewcommand{\Im}{\mathop{\rm Im}\nolimits}

\newcommand{\bver}{\begin{verbatim}}

\newcommand{\defeq}{\stackrel{\rm{def}}{=}}
\def\hto0{\xrightarrow{h\to 0}}

\theoremstyle{plain}
\newtheorem{thm}{Theorem}
\newtheorem{prop}{Proposition}[section]
\newtheorem{cor}{Corollary}
\newtheorem{lem}[prop]{Lemma}

\theoremstyle{definition}

\numberwithin{equation}{section}

\newcommand{\bequ}{\begin{equation}}

\def\bbbone{{\mathchoice {1\mskip-4mu {\rm{l}}} {1\mskip-4mu {\rm{l}}}
{ 1\mskip-4.5mu {\rm{l}}} { 1\mskip-5mu {\rm{l}}}}}

\def\squarebox#1{\hbox to #1{\hfill\vbox to #1{\vfill}}} 
\newcommand{\stopthm}{\hfill\hfill\vbox{\hrule\hbox{\vrule\squarebox 
                 {.667em}\vrule}\hrule}\smallskip} 

\usepackage{epic, eepic}

\title
[Semiclassical $L^p$ estimates]
{Semiclassical $L^p$ estimates}
\author[H. Koch]
{Herbert Koch}
\author[D. Tataru]
{Daniel Tataru}
\author[M. Zworski]
{Maciej Zworski}
\address{ Fachbereich Mathematik, 
	Universit\"at Dortmund,
	D-44221 Dortmund}
\email{koch@math.uni-dortmund.de}
\address{Mathematics Department, University of California \\
Evans Hall, Berkeley, CA 94720, USA}
\email{tataru@math.berkeley.edu}
\email{zworski@math.berkeley.edu}

\begin{document}

\maketitle


\section{Introduction}
\label{in}

\renewcommand\thefootnote{\dag}%

The purpose of this paper is to use semiclassical analysis
to unify and generalize
$ L^p $ estimates on high energy eigenfunctions and spectral clusters. 
In our approach these estimates do not depend on ellipticity and order,
and apply to operators which are selfadjoint only at the principal level.
They are estimates on 
weakly approximate solutions to semiclassical pseudodifferential 
equations. 

To motivate our results let us first 
recall Sogge's 
$ L^p $ estimate \cite{S1} on spectral clusters,
$ \Pi_{[\lambda, \lambda + 1] } $, of 
the Laplace-Beltrami operator, $ - \Delta_g $, 
on a compact Riemannian manifold, $  ( M^n, g ) $: 
\begin{equation} 
\label{eq:sogge}
\Pi_{ [\lambda, \lambda +1] } = {\mathcal O} ( \lambda^{\frac1p} ) 
\; : \; L^2 ( M , d \vol_g) \, \longrightarrow \, 
L^p ( M , d \vol_g) \,, \ \ 
p = \frac{ 2(n + 1 ) }{ n-1} \,,\end{equation}
where 
\[ \Pi_{ I } \defeq  \sum_{ \lambda_j \in I } u_j \otimes \bar u_j \,, \ \
-\Delta_g u_j = 
\lambda_j^2 u_j \,, \ \ \| u_j \|_{ L^2 ( M , d \vol_g ) } = 1\,, \]
and $\{ u_j\} $ form a complete orthonormal set. 

The spectral counting 
remainder estimates of Avakumovi\'c-Levitan-H\"ormander implies 
a bound $ \Pi_{ [ \lambda, \lambda + 1] } = {\mathcal O} ( \lambda^{(n-1)/2} ) 
: L^2 \rightarrow L^\infty $. 
Combining this with \eqref{eq:sogge} and 
the trivial estimate, $ \Pi_{ [ \lambda, \lambda + 1] } = {\mathcal O} ( 
1) : L^2 \rightarrow L^2 $, we obtain optimal $ L^2 \rightarrow L^p $ 
bounds for the spectral cluster operator 
(see the {\color{red} continuous} line in Fig.\ref{fig}).

A similar problem was considered for the harmonic oscillator,
$ -\Delta + |x|^2 $ in $ \RR^n $, 
by  Karadzhov, Thangavelu, and the first two authors 
--- see \cite{KoT} and references given there. In that case, and
for $ n \geq 2$,
 \begin{equation} 
\label{eq:kt}
\Pi_{ [\lambda, \lambda +1] } = \left\{ \begin{array}{ll} 
\ \ \ \ \ \ {\mathcal O} (1) 
& \; : \; L^2 ( \RR^n ) \, \longrightarrow \, 
L^{2n/(n-2)}  ( \RR^n ) \,, \\ 
{\mathcal O} \left( (  \lambda^{ - 1 } \log^{(n+1)/2} \lambda )^{1/(n+3)}
\right) 
& \; : \; L^2 ( \RR^n ) \, \longrightarrow \, 
L^{2(n+3)/(n+1)}  ( \RR^n ) \,, \\ 
\end{array} 
\right. \end{equation}
where now
\[ \Pi_{ I } \defeq  \sum_{ \lambda_j \in I } u_j \otimes \bar u_j \,, \ \
( - \Delta + |x|^2 )  u_j = 
\lambda_j^2 u_j \,, \ \ \| u_j \|_{ L^2 ( \RR^n )  } = 1\,, \]
and again $\{ u_j\} $ form a complete orthonormal set. An interpolated
result  without the logarithmic growth is also valid
(see the {\color{blue} dashed} and dotted
lines in Fig.\ref{fig}). Strichartz estimates 
\cite{KT},\cite{S} lie at the heart of  estimates \eqref{eq:sogge}
and \eqref{eq:kt}. In fact, 
a quick proof of the first estimate in 
\eqref{eq:kt} follows from the pointwise decay of the Schr\"odinger 
propagator and 
the end-point Strichartz estimate of Keel and Tao
\cite{KT}.

A semiclassical point of view -- see \cite{DiSj}, \cite{EZ}, and 
\cite{M} -- allows to put both results in the same setting. 
For compact manifolds we consider the family of 
operators $ -h^2 \Delta_g - 1 $, $ h \sim \lambda^{-1}$, and 
for the harmonic oscillator, $ - h^2 \Delta_y + |y|^2 - 1 $,
where now $ h \simeq \lambda^{-2} $, and $ y = h^{1/2} x $ (see
Example 1 below). 

A natural generalization of the problem 
can then be formulated as follows:
suppose that $ P $ is a semiclassical quantization of a 
classical observable $ p $, that is a
$ P $ is a semiclassical pseudodifferential operator with the 
principal symbol given by $ p $. 
Under what conditions on $ p $ and for what $ \mu ( q ) $ do 
we have
\begin{equation}
\label{eq:gen}
P u = {\mathcal O}_{L^2} ( h ) \,, \ \ \| u \|_{2} = 1 
\; \Longrightarrow \; \| u \|_q = {\mathcal O} ( h^{-\mu ( q ) } ) \, \,  ?
\end{equation}
Here the family of functions $ u = u ( h ) $ is assumed to 
be localized in phase space:
\begin{gather}
\label{eq:locali}
\begin{gathered}
 \exists \; K \Subset \RR^n \,, \  \chi \in \CIc ( \RR^n ) \,,
 \ \text{ independent of $ h $, such that} \\
 \supp u ( h ) \subset K  \ \ \text{and} \  \ \forall \; k \,, 
\ u ( h )  = \chi ( h D) u ( h )  + {\mathcal O}_{H^k} ( h^k ) \,.
\end{gathered}
\end{gather}
An important comment is that the approximate solutions \eqref{eq:gen}
are local, that is, the statement $ P u = {\mathcal O}_{L^2} ( h ) $, 
is invariant under localization in position ($x$) and in momentum ($hD$).


In this introduction we state our results for the 
generalized Schr\"odinger operator,
\begin{equation}
\label{eq:Schr}
   P = - h^2 \Delta_g +  V ( x) \,, \ \ V \in \CI ( \RR^n ; \RR ) \,, 
\ \ \Delta_g = \frac{1}{ \sqrt{\bar g} } \sum_{ i, j = 1}^n \partial_{x_j } 
\sqrt {\bar g  } g^{ ij } \partial_{ x_j } \,, 
\end{equation}
where $ g \defeq ( g^{ij} ( x ) )_{ i,j=1}^n $ is a non-degenerate matrix, and 
$ \bar g \defeq | \det g^{-1}| $. 
The more general results will be presented in the 
sections below. The proofs are based on  semiclassical developments of 
the ideas from \cite{KoT1},\cite{KoT}. 
However, except for the use of basic aspects of semiclassical 
analysis reviewed in Sect.\ref{rsa} and one application of the 
end point Strichartz estimate of Keel and Tao \cite{KT} the paper
is self contained.

\begin{thm}
\label{th:1}
Suppose that $ P $ is given by \eqref{eq:Schr}, $ u = u ( h ) $ 
satisfies \eqref{eq:locali}, and 
\begin{equation}
\label{eq:apr}
  P u = {\mathcal O}_{L^2} ( h ) \,, \ \ \| u \|_{L^2} = 1 \,.
\end{equation}
Then 
\begin{equation}
\label{eq:th13}
\| u \|_p 
= {\mathcal O} ( h^{-\frac1 2} ) \,, \ \ \ p = \frac{ 2 n }{ n-2} \,, 
\ \ n > 2 \,,  
\end{equation}
while for $ n = 2 $
\begin{equation}
\label{eq:th131}
\| u \|_\infty = {\mathcal O} ( ( \log( 1/h ) /h )^{1/2} ) \,.
\end{equation}
If $ V ( x ) \neq 0  $ for $  x \in \supp u $ then 
\begin{equation}
\label{eq:th11}
 \| u \|_p 
= {\mathcal O} ( h^{-\frac1 p} ) \,, \ \ \ p = \frac{ 2 ( n+1) }{ n-1} \,, 
\ \ n \geq 1 \,. 
\end{equation}
\end{thm}

\noindent
{\bf Remark.}
Since we did not assume that $ (g_{ij})_{ 1 \leq i,j \leq n } $ 
is positive definite, but only that it is nondegenerate,  
an example in Sect. \ref{s:nonpr} shows that 
the $ \log ( 1/h ) $ may occur when $ n = 2 $. 
In dimension one, the estimate does not hold as we can 
take $ p ( x, \xi ) = \xi^2 + x^2 $ and $ u ( x ) = h^{-\frac14}
\exp ( -x^2 / (2 h) ) $.

\medskip

The two theorems have some obvious interpolation consequences
which we leave to the reader referring only to Fig.\ref{fig}.
For Schr\"odinger operators we have the following additional 
result which is a generalization of the main result of \cite{KoT},
namely the second estimate in \eqref{eq:kt}.

\begin{thm}
\label{th:new}
Suppose that $ g $ in \eqref{eq:Schr} is positive definite and 
that 
\begin{equation}
\label{eq:dv}  d V ( x ) \neq 0 \,, \ \  x \in \supp u \,. 
\end{equation}
Under the assumptions of Theorem \ref{th:1} we then have 
\begin{equation}
\label{eq:thnew} 
\| u \|_p =  \left\{ \begin{array}{ll}
{\mathcal O} \left( 
h^{\frac16 - \frac 23 n ( \frac{1}{2} - \frac{1}{p}) } 
\right)  & 
\frac{2(n+3)}{n+1} < p \leq \frac{2n}{n-2} \,, \\
\ & \ \\
{\mathcal O} \left(
\log^{\frac{n+1}{2(n+3)} }(1/ h ) h ^{-\frac{n-1}{2(n+3)} }\right) & 
\ \ p = \frac{2(n+3)}{n+1} \,, \\
\ & \ \\
{\mathcal O} \left( 
h^{ - \frac{n-1}2 \left( \frac{1}2 - \frac1p \right) } \right) & 
2 \leq p <  \frac{2(n+3)}{n+1} \,.
\end{array} \right.
\end{equation}
\end{thm}

We do not know if the $ \log ( 1/h ) $ factor
in the estimate \eqref{eq:thnew} is
needed. The optimality of the remaining estimates for the
Hermite operator is discussed in \cite[Sect.5]{KoT}. 
We note that under the assumptions of Theorem \ref{th:new}, we have the 
bound $\| u \|_\infty = {\mathcal O} ( h^{-1/2} ) $ for $ n = 2$.

As a consequence of the two theorems (see Lemma \ref{l:3} below) we have the 
following
\begin{cor}
\label{cor:1}
Suppose that $ P $ is given by \eqref{eq:Schr}, $ u = u ( h ) $ 
satisfies \eqref{eq:locali}, and 
\[  P u = {\mathcal O}_{L^2} ( h ) \,, \ \ \| u \|_{L^2} = 1 \,. \]
Then
\[ \| u \|_\infty = 
 {\mathcal O} ( h^{-\frac{n-1} 2 } ) \,, \ \ n > 2 \,, \ \ 
\| u \|_\infty = {\mathcal O} ( ( \log ( 1/h ) / h )^{\frac 12} ) \,, \ \ 
n = 2 \,. 
\]
If $ V ( x ) \neq 0 $ for $ x \in \supp u $ {\em or} if $ 
(g_{ij}( x) ) $ is positive definite, and $ dV (x ) \neq 0 $ for 
$ x \in \supp u $ then 
\[   \| u \|_\infty =  {\mathcal O} ( h^{-\frac{n-1} 2 } ) \,,  \  \ 
n \geq 1 \,.  \]
\end{cor}

\begin{figure}[htbp]
\begin{center}
\includegraphics[width=6.0in]{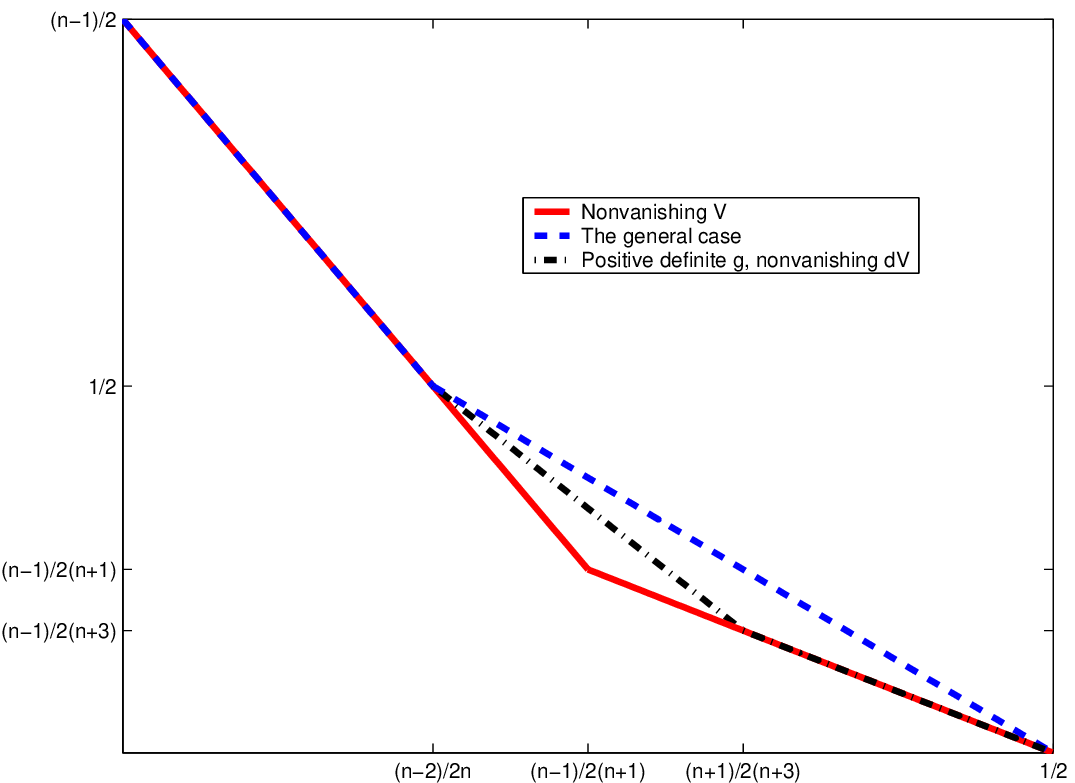}
\end{center}
\caption
{\label{fig} The $x$-axis gives $ 1/p $ and the $ y$-axis 
the power of $ h$ in \eqref{eq:gen}, for $ n \geq 3 $.}
\end{figure}

Returning to the general setting of \eqref{eq:gen}, 
the exponents $ \mu ( q ) $  shown in Fig.\ref{fig} as functions
of $ 1/q $ depend on 
nondegeneracy and curvature properties of the characteristic 
set of $ p $ intersected with the fibers of $ T^*\RR^n $ 
and the support of the localizing function $ \chi $, over the 
support of $ u $.  
In the case of \eqref{eq:gen}
the {\color{red} continuous} lines correspond to estimates 
localized near points at which 
\begin{gather*}
  p ( x_0 , \xi_0 ) = 0 \,,  \ \ d_\xi p ( x_0 , \xi_0 ) \neq 0 \,, \\ 
\{ \xi \; : \; p ( x_0 , \xi) = 0 \} \subset 
T^*_{x_0} \RR^n \ \text{ has a nonvanishing 
second fundamental form at $ \xi_0 $.} \end{gather*}
That case includes Sogge's estimate \eqref{eq:sogge}, or more generally,
the case $ V ( x) \neq 0 $ for Schr\"odinger operators -- see Theorem 
\ref{th:1} and Sect.\ref{Lpn}.

The {\color{blue} dashed} lines correspond to estimates
localized near points at which 
\begin{gather*}
  p ( x_0 , \xi_0 ) = 0 \,,  \ \ d_\xi p ( x_0 , \xi_0 ) = 0 \,, \ \ 
\partial^2_\xi p ( x _0 , \xi_0 ) \ \text{ is nondegenerate.} 
\end{gather*}
This case corresponds to the first estimate in \eqref{eq:kt} -- 
see Theorem \ref{th:1} and Sect.\ref{s:nonpr}. 

The dotted line corresponds to estimates localized 
near points at which 
\[ p ( x_0, \xi_0 ) = 0 \,, \ \ d_\xi p ( x_0, \xi_0 ) = 0 \,, \ \
d_x p ( x_0, \xi_0 ) \neq 0 \,, \ \ \partial_\xi^2 p ( x_0 , \xi_0 ) 
\ \text{is positive definite,} \]
see Theorem \ref{th:new} and
Sect.\ref{isch}. This case corresponds to the second
estimate in \eqref{eq:kt} or the case $ dV \neq 0 $ for 
Schr\"odinger operators.

In this paper we are concerned with smooth symbols only. 
However, similar $L^p $ bounds for Laplacians of $ C^2 $ metrics
were given by Smith in \cite{HS1}, and for $ C^2 $ potentials
in \cite{KoT}. The more robust $ L^\infty $ bounds hold 
for merely $ C^1 $ metrics \cite{HS2}.

Finally, we add that estimates given in Theorems \ref{th:1}, \ref{th:new},
and Corollary \ref{cor:1} are rarely optimal for single eigenfuctions or 
quasimodes -- see \cite{SZ},\cite{STZ} for a discussion and references.
In fact, the problem \eqref{eq:gen} changes dramatically 
when $ P u = {\mathcal O}_{L^2} ( h ) $ is replaced by 
$ P u = o_{L^2 } ( h )$ as the statement can no longer be
localized.


We conclude this introduction with three examples.

\medskip
\noindent
{\bf Example 1.} 
In some cases the scaling allows a transition to some global operators
as in \cite{KoT}.
Suppose that a potential $ W $ satisfies
\[  \ | \partial^\alpha W ( x ) | \leq C_\alpha  \langle x \rangle^{2 - \alpha } \,, \ \ x \in \RR^n \,, \ 
1 \leq i ,j \leq n \,, \]
and
\begin{gather*}  ( - \Delta + W ( x ) - \lambda^2  ) u = {\mathcal O} (1 ) \,, \\ 
\ \forall \; k\,, \  \ \| \langle D \rangle  u \|_2 = {\mathcal O} ( \lambda^{2k} )\,, \ \ 
\supp u \subset \{ x \; :\; \delta \lambda < |x| < \lambda / \delta \} \,.
\end{gather*}
for some $ \delta > 0 $.
Then,
\begin{equation}
\label{eq:5} \| u \|_\infty \leq C \lambda^{ \frac{n-2}2 } \,.
\end{equation}
In fact, put $ h = 1/\lambda^2 $, $ V ( x ) = h W ( x / h^{\frac12}) 
\psi( x/  h^{\frac12} )  -1  $, where $ \psi \in \CIc ( \RR^n \setminus 
\{ 0 \} ) $ satisfies $ \psi ( x  ) u ( x /  h^{\frac12} ) = u ( x/ 
 h^{\frac12} )$. 
A simple rescaling argument and the theorem above give \eqref{eq:5}.
If for instance $ W ( x ) = |x|^2 $ we obtain the natural upper 
bound for the spectral projections:
$$ \bbbone_{ | -\Delta + |x|^2 - \lambda^2 | \leq 1  } = 
{\mathcal O} ( \lambda^{\frac{n-2}2} ) \; : \; L^2 ( \RR^n) \ 
\longrightarrow \ L^\infty ( \RR^n ) \,, \ n \geq 3 \,,$$
and
$$ \bbbone_{ | -\Delta + |x|^2 - \lambda^2 | \leq 1  } = 
{\mathcal O} ( 1 ) \; : \; L^2 ( \RR^n) \ 
\longrightarrow \ L^{\frac{2n}{n-2}} ( \RR^n ) \,, \ n \geq 3 \,.$$

\medskip

\noindent
{\bf Example 2.} 
The $ L^p $ bound in \eqref{eq:th13} is optimal for the ground states of 
\[  - h^2 \Delta + V ( x ) \,, \ \ V ( 0 ) = 0 \,, \ \ 
V'' ( 0 ) \gg 0 \,,  \ \  V \rest_{ x \neq 0 }> 0 \,, 
\ \ \liminf_{ x \rightarrow \infty } V ( x ) 
> 0 \,, \]
that is for $ u ( h ) $ such that
\[  ( - h^2 \Delta + V ( x ) ) u ( h ) = E( h ) u ( h ) \,,  \ \ 
E ( h ) \leq C h \,, \ \ \| u (h ) \|_{2 } = 1\,. \]
see \cite[Chapter 4]{DiSj} and references given there. 

\medskip

\noindent
{\bf Example 3.} Let us consider modes of a damped wave equation on 
a compact Riemannian manifold, $ ( M^n , g ) $,
\[ ( \partial_t^2 + a ( x ) \partial_t - \Delta_g ) u ( t, x ) = 0 \,, \ \ 
u ( t , x) = e^{ - i \tau t } v_\tau ( x ) \,, \ \ 
 \Im \tau \leq 0 \,, \]
$ a \in \CI ( M ; [ 0 , \infty))$, 
see for instance \cite[Sect.5.3]{EZ}. 
Suppose that $ \| v_\tau \|_{L^2 } = 1$, 
Then \eqref{eq:th11} shows that 
\[ \| v_\tau \|_{ p } \leq C |\tau|^{\frac1p} \,, \ 
p = \frac{2 ( n +1)}{n-1} \,,\ 
\ \| v_\tau \|_{\infty } \leq C |\tau|^{(n-1)/2} \,.\]

\medskip

\noindent{\sc Acknowledgments.} We would like to thank
Nicolas Burq, Hart F. Smith III, Christopher D. Sogge, and 
Steve Zelditch for 
helpful discussions related to the topic of this paper. The work of the
first author was supported in part by the Miller Institute at 
the University of California, Berkeley, and that of the 
last two authors by the National Science Foundation grant 
DMS-0354539.

\section{Review of semiclassical analysis}
\label{rsa}

In this section we review basic aspects of semiclassical pseudodifferential
calculus, referring to \cite{DiSj} and \cite{EZ} for details.

We denote by $ T^* \RR^k \simeq \RR^k \times \RR^k $ the cotangent 
bundle of $ \RR^k $. The classical observables are functions of 
position and momentum $ ( x , \xi ) \in   T^* \RR^k $. 
Also, denote by 
$ {\mathcal S} $ and $ {\mathcal S}' $ the space of Schwartz functions
and its dual respectively, and 
suppose that $ a \in {\mathcal S}( T^*\RR^k ) $ and . Then the {\em left 
semiclassical quantization} of $ a $ is the operator 
$ a ( x, h D ) :  {\mathcal S}' ( \RR^k ) \longrightarrow 
{\mathcal S} ( \RR^k ) $ densely defined by 
\[ a ( x , h D) u ( x )  \defeq \frac{1}{(2 \pi h)^{k}} \int e^{\frac i h \langle x - y, \xi \rangle } a ( x , \xi ) u ( y ) dy  d \xi \,, \ \ 
u \in { \mathcal S} ( \RR^k ) \,. \]
In a few places it will be convenient to use the Weyl quantization,
\[  a^w ( x , h D) u ( x )  \defeq 
\frac{1}{(2 \pi h)^{k}} \int e^{\frac i h \langle x - y, \xi \rangle } 
a \left( \frac{x+y}2 , \xi \right) u ( y ) dy  d \xi \,, \ \ 
u \in { \mathcal S} ( \RR^k ) \,. \]
One of its advantages is the selfadjointness of $ a^w ( x , h D ) $ 
for real values $ a$'s.

This definition can be extended to a large class of observables.
A function, $ m : T^*\RR^k \longrightarrow [0, \infty ) $ is called
an {\em order function} if for all $ ( x, \xi ) \,, \ ( y ,\eta) 
\in T^* \RR^k $, 
\[ m ( x ,\xi ) \leq C ( 1 + | x - y | + | \xi - \eta| )^N m ( y , 
\eta ) \,, \]
for some fixed $ C $ and $ N $. We say that $ a \in \CI ( \RR^k ) $ 
is a symbol in class $ S ( m ) $ if 
\[ | \partial^{\alpha}_{ x , \xi } a  (x , \xi ) | \leq 
C_ \alpha m ( x , \xi ) \,, \ \ \alpha \in \NN^{2k} \,. \]
Unless specifically stated, we always allow the symbols to depend on 
$ h $.
The continuous map 
\[ {\mathcal S} ( T^* \RR^k ) \ni a \longmapsto a ( x , h D ) 
\in {\mathcal L} (   {\mathcal S} ( \RR^k )
,  {\mathcal S} ( \RR^k ) ) \,, \]
extends to a continuous map 
\[ S( m ) \ni a \longmapsto a ( x , h D ) 
\in {\mathcal L} (   {\mathcal S} ( \RR^k ) ,  {\mathcal S} ( \RR^k ) ) \
\,, \] 
which satisfies the following fundamental composition property:
if $ m_j $, $ j =1, 2 $ are two order functions, and $ a_j \in S( m_j ) $,
$ j = 1 , 2 $, then 
\begin{equation}
\label{eq:comp} 
a_1 ( x , hD ) a_2 ( x , h D) = b ( x , h D ) \,, \ \ 
b \in S ( m_1 m_2 ) \,.
\end{equation}
Moreover we have an asymptotic formula for $ b ( x , \xi ) $ given by 
\begin{equation}
\label{eq:asymp}
b ( x , \xi ) \sim \sum_{ \alpha \in \NN^k } 
\frac1{\alpha !} \partial_\xi ^\alpha a_1 ( x , \xi ) ( hD_x)^\alpha 
a_2 ( x , \xi ) \,. \end{equation}
We also have the mapping property:
\begin{equation}
\label{eq:mapp}
a \in S ( 1 ) \; \Longrightarrow \; a ( x , h D) = {\mathcal O}( 1) 
\; : \; L^2 ( \RR^k ) \longrightarrow L^2 ( \RR^k ) \,. 
\end{equation}

Suppose that $ a \in S ( 1 ) $
and that $ |a ( x , \xi )| \geq 1/C $ for all $ ( x , \xi ) \in 
T^* \RR^k $. Then 
\[  a ( x, h D) ^{-1} \; : \; L^ 2 ( \RR^k ) \; \longrightarrow 
\; L^2 ( \RR^k ) \,,\]
exists if $ h $ is small enough. In fact, by our hypothesis
$ c ( x , \xi ) \defeq 1/a ( x , \xi ) \in S ( 1 ) $, and by 
\eqref{eq:comp} and \eqref{eq:asymp}, 
\[ a ( x , h D ) c ( x , h D ) = I + h r ( x , h D ) \,, \ \ 
r \in S ( 1) \,.\]
By \eqref{eq:mapp}, $ r ( x, h D ) = {\mathcal O} ( 1 ) 
: L^2 \rightarrow L^2 $, and hence $ I + h r ( x , h D) $ is 
invertible for $ h $ small enough. This gives 
$ a ( x , h D) ^{-1} = c ( x, h D ) ( I + h r ( x , h D) )^{-1}$.
The use of semiclassical Beals's Lemma \cite[Proposition 8.3]{DiSj},
\cite[Theorem 8.9]{EZ}, shows more: $ a ( x, h D)^{-1} = b( x , hD)
$, $ b \in S ( 1 ) $.

In this note we will also need a microlocal version of this 
result:
\begin{lem}
\label{l:1}
 Suppose that $ \chi \in S ( 1 ) $, $ m $ is an 
order function,  and that $ a \in 
S ( m ) $ satisfies
$ | a ( x , \xi ) | \geq m ( x, \xi ) / C  $ 
for $ ( x , \xi ) \in \supp \chi $.
Then there exists $ b \in S ( 1/m ) $ such that 
\begin{equation}
\label{eq:l1}
\begin{split}
&  b ( x , h D ) a ( x , h D) \chi ( x , h D) = \chi ( x , h D) + 
{\mathcal O}_{ L^2 \rightarrow L^2} ( h^\infty ) \,,\\
&  a ( x , h D ) b ( x , h D) \chi ( x , h D) = \chi ( x , h D) + 
{\mathcal O}_{ L^2 \rightarrow L^2  } ( h^\infty ) \,.
\end{split}
\end{equation}
When $ \chi \in \CIc (T^*\RR^n)  $ we can replace 
$ {\mathcal O}_{ L^2 \rightarrow L^2  } ( h^\infty ) $ by 
$ {\mathcal O}_{ {\mathcal S}' \rightarrow {\mathcal S} } ( h^\infty ) $.
\end{lem}
\begin{proof}
We give the proof in the case of $ \chi \in \CIc $ and 
we first note that $ a \chi \in S ( ( 1 + |x| + |\xi|)^{-M}  ) $ for
any $ M $. We then inductively construct $ b_j \in S( 1) $ such that
\[ \left( \sum_{j=0}^N h^j b_j ( x, h D ) \right) a ( x , h D) \chi ( x, h D) = 
\chi ( x , h D ) + h^N r_N ( x , h D ) \,, \] 
$ r_N \in  S ( ( 1 + |x| + |\xi|)^{-M}  ) $,
for any $ M $. The symbol $ b \in S ( 1) $ satisfying 
\[ b ( x, \xi) \sim \sum_{j=0}^\infty b_j ( x, \xi) \]
gives \eqref{eq:l1}.
\end{proof}

The next lemma provides basic semiclassical $L^p$ estimates:
\begin{lem}
\label{l:2}
Suppose that $ a \in {\mathcal S} ( T^* \RR^k ) $. Then for $ 1 \leq q \leq
p  \leq \infty  $,
\begin{equation}
\label{eq:l2}
 a ( x , h D) = {\mathcal O} ( h^{k \left( 1/ p -  1/ q \right) } ) \; : \; L^q ( \RR^k ) \; \longrightarrow \; L^p ( \RR^k ) \,. 
\end{equation}
\end{lem}
\begin{proof}
We first recall that 
\[  a ( x , h D) u( x )  = h^{-k} 
\int K ( x, ( x - y ) / h ) u ( y ) dy \,,\]
where 
\[ K ( x , z ) \defeq \frac{1}{(2\pi )^k }\int a ( x , \xi ) e^{ i \langle
\xi , z \rangle } d\xi \,.\]
In particular $ K \in {\mathcal S} ( \RR^k \times \RR^k ) $, and 
$ | K ( x , z ) | \leq C_N ( 1 + | z|)^{-N} $ for any $ N $ with 
$ C_N $ independent of $ x $.
This means that 
\[ \| a ( x , h D) u \|_{ L^p} \leq C_N h^{-k} \| ( 1 + |\bullet/h| )^{-N}
* u \|_{p } \,.\]
The Young inequality, 
\begin{equation}
\label{eq:Young} \| f * u \|_{ L^p } \leq \| f \|_{r} \| u \|_{q} \,, \ \
1 \leq p,q,r \leq \infty \,, \ \ \frac 1 p + 1 = \frac 1 r + \frac 1 q \,,
\end{equation}
and the calculation 
\[ h^{-k}  \| ( 1 + |\bullet/h| )^{-N} \|_{ L^r } = C h^{-k} h^{k/r} = 
C h^{ k ( 1/p - 1/q) } \,,\]
give \eqref{eq:l2}.
\end{proof}

A microlocal version of the localization assumption \eqref{eq:locali}
is given as follows
\begin{equation}
\label{eq:loc} \exists \; \chi \in \CIc ( T^* \RR^k ) \,, \ N \geq 0 \,, \ \ 
u ( h ) = \chi ( x , h D) u ( h ) + {\mathcal O}_{\mathcal {S} }
 ( h^\infty ) 
\,,  \ \ \| u ( h ) \|_{2 } = {\mathcal O} ( h^{-N} )  \,.
\end{equation}
The  bound  in $L^2 $ is needed as otherwise the statement
$ {\mathcal O}_{ \mathcal S } ( h^\infty ) $ has no meaning, in 
view of scaling. We also need it to guarantee that the residual terms 
in the semiclassical calculus give $ {\mathcal O}(h^\infty ) $ bounds
when applied to $ u ( h ) $.  Except in Theorem \ref{th:ell} we 
can simply assume that $ \| u ( h ) \|_{L^2 } = 1 $.

This assumption combined with Lemma \ref{l:2} has the following
consequence which is a semiclassical version of Sobolev embedding.
In fact, it is equivalent to Sobolev embedding for functions 
localized in frequency to a dyadic corona.

\begin{lem}
\label{l:3}
Suppose that a family $ u = u ( h ) $ satisfies \eqref{eq:loc}.
Then for any $ 1 \leq q \leq p \leq \infty $,
\begin{equation}
\label{eq:l3}
\| u \|_{ L^p } \leq C h^{ k ( 1/p - 1/q ) } \| u \|_{ L^q } 
+ {\mathcal O} ( h^\infty ) \,.
\end{equation}
\end{lem}
\begin{proof}
The estimates for $ \| \chi ( x, h D_x)  u \|_{p } $ follows 
from Lemma \ref{l:2} and 
\[ \| ( 1 - \chi ) ( x , h D )   u \|_{q } = O ( h^\infty ) \,, \]
from \eqref{eq:loc}
\end{proof}

As an application of Lemmas \ref{l:1} and \ref{l:3} we state
the following {\em elliptic} semiclassical $ L^p $ estimate.
It shows that to obtain general estimates in the remaining
sections we can assume that $ u $ is localized to a neighbourhood
of a characteristic point of $ P$.

\begin{thm}
\label{th:ell}
Suppose that $ u $ satisfies the localization condition 
\eqref{eq:loc} and that 
\[ P u = {\mathcal O}_{L^2} ( h ) \,, \ \  | p ( x , \xi ) | \geq 1/C 
\,, \ \ ( x, \xi ) \in \supp \chi \,.\]
Then 
\[  \| u \|_{ p } = {\mathcal O} ( h^{ 1 - n ( 1/2-1/p) } ) \,.
\] 
\end{thm}

The next lemma is a global semiclassical version of 
a Sobolev embedding estimate
(see for instance \cite[Theorem 4.5.13]{H1}):

\begin{lem}
\label{l:Sob}
Suppose that $ \Omega_1 $ and $ \Omega_2 $
have properties stated in Lemma \ref{l:ell}. Then for 
$ u \in \CI ( \RR^n ) $
\[ \| u \|_{L^p ( \Omega_1 ) } \leq C_1 h^{-n (1/2-1/p) } \sum_{ |\alpha | 
\leq m } \| ( h D)^\alpha u \|_{ L^2 ( \Omega_2 ) } \,, \ \ 
\frac{1}2 - \frac m n \leq \frac1 p  \leq \frac 12 \,, \ \ p < \infty \,.
\]
When $ u \in \CIc ( \RR^n) $, and 
$ 1/2 - m/n = 1/p $, $ p < \infty $,  we can replace $ | \alpha | \leq m $
in the sum by $ | \alpha |= m  $.
\end{lem}
\begin{proof}
We can assume that $ u \in \CIc ( \Omega_2 ) 
$ and then can consider $ \Omega_1 =
\Omega_2 = \RR^n $. In that case the estimate 
with $ h = 1 $ is a standard Sobolev inequality. Applying it
to $ v_h ( x ) =
u ( h x  )$ gives the lemma: $ (h D_x)^{\alpha } u
= D_x^\alpha v_h $,
\[ \| v_h \|_{ H^m ( \RR^n ) } = h^{-\frac{n}2}  \sum_{
|\alpha| \leq m } \| ( h D_\alpha)  u \|_{L^2 ( \RR^n) }
\,, \ \
\| v_h \|_{ L^p ( \RR^n )}  = h^{ -\frac{n}{p} } \| u \|_{L^p ( \RR^n ) } \,.\]
\end{proof}

For future reference we state also another basic fact. Let
\[  {\mathcal F}_{ h } v (  \xi ) \defeq 
 \frac{1}{ (2 \pi  h)^{n/2}} \int_{\RR^n}  v ( x ) e^{ \frac i h \langle 
x , \xi \rangle } dx \,, \]
be the semiclassical Fourier transform, normalized to be unitary on 
$ L^2 ( \RR^n ) $. The semiclassical Sobolev spaces are defined 
using the following norm 
\[ \| u \|_{ H^s_h ( \RR^n ) }^2 \defeq 
 \int_{\RR^n} ( 1 + |\xi|^2 )^{s} | {\mathcal F}_h v ( \xi ) |^2 
d \xi \,.\]
If $ s $ is a nonegative integer then clearly
\[  \| u \|_{ H^s_h ( \RR^n ) } \simeq 
\sum_{ | \alpha | \leq s } \| ( h D)^\alpha u \|_2  \,.\]

\begin{lem}
\label{l:Ber}
For $ s > n/2 $ we have 
\[ \| u \|_\infty \leq C h^{-n/2} \| u \|_{H^s_h ( \RR^n ) }  \,.\]
\end{lem}
\begin{proof}
We follow the usual procedure keeping track of the 
parameter $ h $: 
\[ \begin{split}   \| u \|_\infty^2 & \leq 
\frac { 1 } {( 2 \pi h) ^{n}  } 
\left( \int_{\RR^n }
| {\mathcal F}_{h } u ( \xi ) | d \xi \right)^2 
\\ & 
 \leq 
\frac { 1 } { ( 2 \pi h)^{n} } \int_{\RR^n } ( 1 + |\xi|^2 )^{-s} 
d \xi 
\int_{\RR^n} ( 1 + |\xi|^2) )^s
|  {\mathcal F}_{h } u ( \xi ) |^2 d \xi =  \frac{ C }{ h^{n} } 
\| u \|_{ H^s _h ( \RR^n ) }^2 \,. 
  \end{split} 
\]
\end{proof}

Finally, we state without proof a semiclassical version of standard 
elliptic estimates (see for instance \cite[Theorem 17.1.3]{Hor2}):

\begin{lem}
\label{l:ell}
Suppose that a differential operator, $ P ( h ) = \sum_{ |\alpha | 
\leq m } a_\alpha ( x , h ) ( h D_x )^\alpha $, satisfies,
\begin{equation}
\label{eq:const} \forall \, | \alpha | \leq m \,, \ 
\beta \in \NN^n \,, \ \  \partial_x^\beta a_{\alpha} ( x , h ) 
= {\mathcal O}( 1) \,, \ \   | \sum_{ |\alpha | = m } a_{\alpha}
( x , h ) \xi^\alpha | \geq | \xi |^m/C  \,, \ \ C > 0 \,,
\end{equation}
uniformly for $ x \in K$, for any $ K \Subset \RR^n $.
Then for any bounded 
open sets $ \Omega_1 $, $ \Omega_2 $, $ \overline \Omega_1 \Subset \Omega_2$,
and $ u \in \CI ( \RR^n ) $, we have 
\[ \sum_{ |\alpha | \leq m } \| ( h D)^\alpha u \|_{L^2 ( \Omega_1 ) } 
\leq C_0 \left( \| P ( h ) u \|_{ L^2 ( \Omega_2 ) } + 
\| u \|_{ L^2 ( \Omega_2) } \right) \,, \]
where $ C_0 $ depends only on constants in \eqref{eq:const} for $ K =
\overline \Omega_2 $, $ \Omega_2$, and $ \Omega_1 $.
\end{lem}

\section{$L^\infty$ estimates in the principal type case}

In this section we prove $ L^\infty $ bounds under a principal
type assumption. We remark that this assumption is always satisfied
in the case of the Laplacian on a Riemannian manifold for which 
$ p ( x , \xi ) = \sum g^{ij}( x)  \xi_i \xi_j -1 $. 
The simple direct proof implies, rather than uses, 
the optimal upper bound on the 
number of eigenvalues of an elliptic operator in an interval of
size $ h $ -- see Corollary \ref{cor:2} at the end of this section.

\begin{thm}
\label{th:8.5}
Let 
$ m = m ( x , \xi ) $ an order function, and
let $ u ( h ) \in L^2 ( \RR^n ) $ satisfy
the frequency localization condition \eqref{eq:loc}.
Suppose that $ p \in S ( m ) $ is real valued, 
and that
\begin{equation}
\label{eq:xinod}
 p ( x , \xi ) = 0 \,, \ \ ( x, \xi ) \in \supp  \chi  \ \Longrightarrow 
\partial_\xi p ( x , \xi ) \neq 0 \,. \end{equation}
Then 
\begin{equation}
\label{eq:Linft2} 
\| u ( h ) \|_{ L^\infty } 
\leq C h^{-(n-1)/2 } \left(  \| u ( h ) \|_{ L^2} + 
\frac 1 h  \| p ( x , h D ) u ( h ) \|_{ L^2 } \right) \,. 
\end{equation}
\end{thm}

\bigskip

\noindent {\bf Remark.} 
The bound given in Theorem \ref{th:8.5} is already optimal in the
simplest case in which the assumptions are satisfied: 
$p ( x , \xi ) = \xi_1 $.
Indeed, write $ x = ( x_1, x') $ and let $ \chi_1 \in \CIc ( \RR ) $,
and $ \chi \in \CIc ( \RR^{n-1} ) $. Then 
$$ u ( h ) := h^{ -(n-1)/2} \chi_1 ( x_1 ) \chi( x'/h ) $$ 
satisfies 
\[ p ( x , h D)  u ( h ) =  hD_{x_1} u( h ) = O_{L^2} ( h ) \,, \ \ 
\| u ( h ) \|_{2} = O ( 1 ) \,, \]
and for any non-trivial choices of $\chi_1 $ and $ \chi $, 
\[ \| u ( h ) \|_{\infty } \simeq h^{-(n-1)/2} \,. \]
The condition \eqref{eq:xinod} is in general necessary as shown by 
another simple example. Let $ p ( x , \xi ) = x_1 $, and
$$ u (h ) = h^{-n/2} \chi_1 ( x_1/h ) \chi ( x'/h ) \,. $$
 Then 
\[ P ( h ) u ( h ) =  h h^{-n/2} ( t \chi_1 ( t) )|_{t=x_1/h} 
\chi( x'/h ) = O_{L^2} ( h ) \,, \ \ \| u ( h ) \|_{2} = O ( 1 ) \,, \]
and 
\[ \| u ( h ) \|_{\infty } \simeq h^{-n/2} \,, \]
which is the general bound of Lemma \ref{l:3}.

\bigskip

\noindent
{\em Proof of Theorem \ref{th:8.5}:} First we observe that
we can assume that $ u ( h ) $ is compactly supported.
We also note that the estimate
hypothesis on $ u ( h ) $ is local in phase space: if $ \chi \in 
\CIc ( T^* \RR^k ) $ then, normalizing to $ \| u ( h ) \|_{2 }
= 1$,  
\[ \begin{split}
p( x, hD) \chi^w ( x, h D )  u ( h ) & = \chi^w ( x , h D) p( x, hD) u ( h ) 
+ [ p( x, hD) , \chi^w ( x , h D ) ] u ( h )\\
&  = {\mathcal O} (1) ( h \| u \|_{2} + \| p ( x , h D ) u \|_{2} ) \,, \end{split} \]

 Hence it is enough to prove the theorem for $ u ( h ) $ replaced by 
$ \chi^w u ( h ) $, where $ \chi $ is supported near a given point in 
$ K $ as a partition of unity  argument will then 
give the bound on $ u ( h ) $. A partition of unity, in this case, 
means a set of functions,
$$ \{ \chi_j \}_{j=0}^N \subset \CIc ( T^* \RR^n) \,, 
$$ such that
\begin{gather}
\label{eq:partun}
\begin{gathered}
\sum_{j=1}^N \chi_j ( x, \xi ) =  \chi_0 ( x, \xi  )  \,, 
\ \ \supp \chi_j \subset U_j \,, \ \ \supp \chi_0 \subset U_0 \defeq 
\bigcup_{j=1}^N 
U_j\,, 
\end{gathered}
\end{gather}
where $ U_0 $ is a neighbourhood of $ \supp \chi $, a compact set, 
in which \eqref{eq:xinod} holds.

Suppose that $ p \neq 0 $ on the support of $ \chi $. We can 
quote Theorem \ref{th:ell} but for the reader's convenience 
present an argument. From the ellipticity and Lemma \ref{l:1} 
we see that
$ p( x, hD) \chi^w u ( h ) = {\mathcal O}_{L^2} ( h ) $ implies that
$ \chi^w u ( h ) = {\mathcal O}_{L^2} ( h ) $. Lemma \ref{l:3}
then shows that 
$$ \| \chi^w u ( h ) \|_{\infty} \leq C h h^{-n/2} \leq C h^{-(n-1)/2 } 
\,.$$

Now suppose that $ p $ vanishes in the support of $ \chi $. 
By applying a linear change of variables we can assume that $ p_{\xi_1} 
\neq 0 $ there. The implicit function theorem 
shows that 
\begin{equation}
\label{eq:pel}
 p ( x , \xi ) = e ( x , \xi ) ( \xi_1 - a ( x , \xi') ) \,, \ \ 
\xi = ( \xi_1 , \xi' ) \,, \  \ e ( x , \xi ) > 0  \,, 
\end{equation}
holds in a neighbourhood of $ \supp \chi $. We extend $ e $ 
arbitrarily to $ e \in S $, $ e \geq 1/C $, and $ a ( x , \xi') $
to a real valued $ a ( x, \xi' ) \in S $. The pseudodifferential 
calculus shows that 
\[ \begin{split}
e^w ( x , h D) ( h D_{x_1} - a ( x , hD_{x'} ) ) ( \chi^w u ( h ) ) 
& = p( x, hD) ( \chi^w u ( h )) + {O}_{L^2} ( h ) \\
& =  {O}_{L^2} ( h )  \,, \end{split} \]
and since $ e^w $ is elliptic,
\begin{equation}
\label{eq:eel}
  ( h D_{x_1} - a ( x , hD_{x'} ) ) ( \chi^w u ( h ) ) = O_{L^2} ( h ) 
\,. \end{equation}

The proof will be completed if we show that 
\begin{equation}
\label{eq:redL2} 
   \| ( \chi^w u) ( x_1 , \bullet) \|_{ L^2 ( \RR^{n-1} ) }
= O( 1)  \,, 
\end{equation}
and for that  we need another elementary
\begin{lem}
\label{l:simple}
Suppose that $ a \in S ( \RR \times T^* \RR^k ) $ is real valued, and that
\begin{gather*}
( h D_t + a^w ( t , x, hD_x ) ) u ( t , x) = f ( t , x) \,, \ \ u ( 0 , x ) = 
u_0 ( x ) \,, \\ f \in L^2 ( \RR \times \RR^k ) \,, \ \
u_0  \in L^2 ( \RR^k ) \,. 
\end{gather*}
Then 
\begin{equation}
\label{eq:sqrtt}  \| u ( t , \bullet ) \|_{ L^2 ( \RR^k ) } \leq 
\frac{\sqrt{t}}{h} \| f \|_{L^2 ( \RR \times \RR^k ) }  + 
\| u_0 \|_{ L^2 ( \RR^k ) } \,. \end{equation}
\end{lem}
\noindent
\begin{proof} Since $ a^w ( t , x , h D) $ is family of bounded 
operators on $ L^2 ( \RR^k ) $ existence of solutions follows from 
existence theory for (linear) ordinary differential equations in $ t $. 
Suppose first that $ f \equiv  0 $. Then 
\[ \begin{split} 
\frac12 \frac{d}{dt} \| u ( t ) \|_{L^2( \RR^k)  }^2 & = 
\Re \langle \partial_t u (t ) , u ( t ) \rangle_{L^2 ( \RR^k ) } \\
& = \frac{1}{h} \Re \langle i a^w ( x, h D ) u ( t ) , u ( t ) \rangle = 0 
 \,. \end{split} \]
Thus, if we put $ E ( t ) u_0 := u ( t) $,
\[  \| E ( t ) u_ 0 \|_{ L^2 ( \RR^k ) } = \| u_0 \|_{L^2 ( \RR^k ) } \,.  \]
If $ f \neq 0 $,  Duhamel's formula gives
\[ u ( t ) = E( t) u_0 + \frac{i}{h} \int_0^t E ( t - s ) f ( s ) ds \,,\]
and hence
\[ \| u ( t ) \|_{ L^2 ( \RR^k ) } \leq \| u_0 \|_{L^2 ( \RR^k ) }
+ \int_0^t \| f ( s ) \|_{ L^2 ( \RR^k ) } \,. \]
The estimate \eqref{eq:sqrtt} is an immediate consequence.
\end{proof}

The estimate \eqref{eq:redL2} is immediate from the lemma and \eqref{eq:eel}. 
We now apply Lemma \ref{l:3} in $ x' $ variables only, that is 
with $ k = n-1$. That is allowed since we clearly have 
\[  \| ( 1 - \psi ( h D' ) ) \chi^w u ( h ) ( x_1 , \bullet ) \|_{L^2 ( 
\RR^{n-1} ) } = O ( h^\infty ) \,, \]
uniformly in $ x_1 $.
\stopthm

As an application we give a proof of a well known result about the 
density of eigenvalues near a nondegenerate energy level -- 
see \cite[Chapter 4]{Iv} for a full discussion. For
simplicity we assume that our operator is defined on a compact 
manifold $ X $ -- see \cite[Appendix D]{EZ} for an introduction to 
semiclassical analysis on manifolds. The symbol classes are now
defined as
\[  S^{m,k} ( T^* X ) = \{ a \in \CI ( T^*X ) \; : \; 
| \partial_x^\alpha \partial_\xi^\beta a | \leq h^{-k} C_{\alpha 
\beta } (  1+ |\xi|)^{m-|\beta | } \} \,,\]
with corresponding operators denoted by $ \Psi^{m,k} ( X , \Omega_{\frac12} X 
 ) $, where to avoid a choice of a density we act on half densities
on $ X $ (see \cite[Sect.8.1]{EZ}).
The principal symbol of $ P \in \Psi^{m,k} ( X , \Omega_{\frac12} ) $ is then
defined in $ S^{m,k}/S^{m-1,k-1} ( T^* X ) $. The example to 
keep in mind is of course
\[ P = -h^2 \Delta - 1 \in \Psi^{2,0} ( X ) \,.\]

\begin{cor}
\label{cor:2}
Let $ P \in \Psi^{m,0} ( X , \Omega_{\frac12} X ) $ 
be a semiclassical selfadjoint
pseudodifferential 
operator on a compact $ n$ dimensional 
manifold $ X $ with a {\em real} principal symbol 
$ p 
\in S^{m,0} ( T^* X ) $ (well defined modulo $S^{m-1,-1} ( T^* X ) $)
satisfying 
\[   | p ( x , \xi )| \geq ( 1 + |\xi|)^{m}/C - C \,, \ \ ( x , \xi ) 
\in T^*X \,.\]
Let $ \Spec ( P ) \subset \RR  $ 
be the spectrum of $ P $ which 
is a discrete set. If 
\[ p ( x , \xi ) = E \in \RR  \implies d_\xi p ( x, \xi ) \neq 0 \,, \]
then 
\[ | [ E -  h ,E + h ] \cap \Spec ( P ) | = {\mathcal O} ( h^{1-n} )
\,. \]
\end{cor}
\begin{proof} We reverse the standard argument for obtaining
$L^\infty $ bounds from remainder estimates for the spectral 
projection -- see \cite{S1}. Under the assumptions on $ P $,
the resolvent $ ( P - z )^{-1} $ is compact for $ z \notin 
\RR $ (for instance using Lemma \ref{l:1} with $ m ( x, \xi) 
= \langle \xi \rangle^m $). Hence the spectrum consists of
isolated eigenvalues, $ \lambda $,  with smooth eigenfuctions
half densities, $ \phi_\lambda $. We define the spectral 
projection, 
\[ \Pi_h ( x , y ) = \sum_{ |\lambda - E | \leq h } \phi_\lambda ( x )
\overline {\phi_\lambda ( y ) } \,. \]
Theorem \ref{th:8.5} shows that 
\[ \Pi_h = {\mathcal O} ( h^{-(n-1)/2} ) \; : \; L^2 ( X , \Omega_{\frac12} X ) 
\rightarrow L^\infty ( X ) \,. \]
Here we chose a trivialization of the half-density bundle which identified
half densities with functions, allowing a map into $ L^\infty $.
Hence, 
\[  \Pi_h ( x , x ) = \int_{X_y}  \Pi_h ( x , y ) \Pi_h ( y , x ) = 
F_h ( x) |dx| \,, \ \ | F_h ( x ) | \leq \| \Pi_h \|_{ L^2 \rightarrow L^\infty }^2  
\leq C   h^{-n+1} \,, \]
and 
\[ \begin{split} | [ E -  h ,E + h ] \cap \Spec ( P ) | 
&  = \int_X \Pi_h ( x , x ) = \int_X  F_h ( x )  |dx|  \\
&  \leq \vol(X) \| F_h \|_\infty = 
{\mathcal O} ( h ^{-n+1 } )  \,. \end{split} \]
Here the volume was computed using the same trivialization of the 
density bundle.
\end{proof}

The same proof can be applied in other situations in which 
we have precise $ L^\infty $ bounds, for instance under the assumptions
of Theorems \ref{th:1}, \ref{th:nonpr}, $n > 2$. That however does not
add anything new to the results of Ivrii \cite{Iv}. 
Brummelhuis-Paul-Uribe \cite{BPU} obtained precise asymptotics
when the critical set of $ p $ has a nice structure and that paper 
can be used to construct operators for which $ \log ( 1/h ) $ 
appears in $ L^\infty $ bounds. 
However, both  
references suggest that the $ \log (1/h ) $ term in Theorem \ref{th:1}
when $ n = 2 $ does not occur for Schr\"odinger operators.

Finally we remark that in the case of nonselfadjoint operators
$ L^\infty $ estimates do not seem to give bounds on the 
number of eigenvalues in small regions -- see \cite{SjZw04} for
a discussion of such estimates and references in the context of resonances.

\section{Semiclassical Strichartz estimates}

To prove Theorems \ref{th:1} and \ref{th:new}, or rather 
their more general versions in Sections \ref{Lpn}, \ref{s:nonpr},
and \ref{isch}, we use Strichartz estimates. Unlike the 
$ L^\infty $ bound of the previous section which involved
an energy estimate only they rely on
the nondegeneracy of $ \partial_\xi^2 p $. 

Semiclassical Strichartz estimates for the Schr\"odinger 
propagator of $ P = -h^2 \Delta_g - 1$
appeared explicitely in the work of Burq, G\'erard, and Tzvetkov \cite{BGT}
who used them to prove existence results for non-linear Schr\"odinger
equations on two and three dimensional compact manifolds. 
A more robust phase space representation of Schr\"odinger propagators
applicable to a wider range of operators is given in 
\cite{KoT1} and \cite{Tat}. 
We refer to these papers for pointers to the vast literature on
Strichartz estimates and their applications. 

Here we give a consequence of the well known
parametrix construction recalled in
Proposition \ref{p:osc} and of the abstract Strichartz
estimates of \cite{KT}. For the reader's convenience we first
recall the abstract Strichartz estimate, slightly modified 
for the semiclassical application:
\begin{prop}
\label{t:KT}
Let $ ( X, {\mathcal M} , dm ) $ be a $ \sigma$-finite measure space, and 
let 
$$ U \in L^\infty (  \RR , 
 {\mathcal B} ( L^2 ( X , dm ) ) $$
satisfy
\begin{gather}
\label{eq:kt1}
\begin{gathered}
\| U ( t ) \|_{ {\mathcal B}  ( L^2 ( X ) )} \leq A  \,,  \ \ 
t \in \RR \,, \\
\| U ( t )  U ( s )^* f \|_{L^\infty ( X , \mu ) } \leq 
A h^{-\mu} (| t - s | +h)^{-\sigma}\| f \|_{ L^1 ( X , dm ) }  \,, \ \ t , s \in \RR \,,
\end{gathered}
\end{gather} 
where  $ A, \sigma >0 $, $ \mu \geq 0 $ are fixed. 

The for every pair $ p , q  $ satisfying 
\[ \frac{2}{p} + \frac{2 \sigma }{q } = \sigma \,, \ \  2 \leq p \leq \infty 
\,, \ \ 1 \leq q \leq \infty \,, \ \ 
( p , q ) \neq ( 2, \infty ) \,, 
\]
we have 
\begin{equation}
\label{eq:kt2}
\left( \int_{\RR} \|U ( t ) f \|_{L^q ( X, dm) } 
^p dt \right)^{\frac1p} \leq B h^{ -\frac{ \mu } {p
\sigma} } \| f \|_{L^2 ( X , dm ) } \,.
\end{equation}
When $ ( p , q ) = ( 2, \infty ) $, and $ \mu = 2$, 
we have the same estimates with 
the $ h$ dependent constant replaced by $ 
(\log ( 1/h )/h)^{1/2} $.
\end{prop}
To explain the logarithmic correction term for $ ( p ,q ) = 
( 2, \infty ) $, that is $ \sigma = 1 $, we recall the proof in in that case
referring the reader to \cite{KT} for a complete argument. We
also remark that in \eqref{eq:kt1} $ ( | t - s | + h )^{-\sigma} $
can be replaced by $ | t - s|^{-\sigma } $ except for the case of
$ ( 2, \infty ) $.

\noindent
{\em Proof of the case $ \sigma = 1 $, $ p = 2$:}
The estimate we want reads
\begin{equation*}
\| U ( t) f \|_{ L^2 (\RR_t  , L^\infty ( X ))  } \leq  
B \| f \|_{ L^2 ( X ) }
\,.
\end{equation*}
This is equivalent to 
\[ \int_{ \RR \times X} U ( t ) f ( x ) \; G ( t, x) \; dm ( x ) dt 
\leq \| f \|_{ L^2 ( X ) } \| G \|_{ L^{2} ( \RR , L^1 ( X ) ) } \,, \]
for all $ G \in L^{2} ( \RR , L^\infty ( X ) ) $, and that in turn means that
\[ \| \int_\RR  U ( t )^* G ( t )  dt \|_{L^2 (X ) } 
\leq C \| G \|_{ L^{2} ( \RR , L^1 ( X ) )  } \,, \]
or in other words that
\begin{equation}
\label{eq:tts}
 T :   L^{2} ( \RR , L^\infty ( X ) ) \longrightarrow  
L^2 ( X ) \,, \ \  T G ( x ) := \int_\RR U ( t ) ^* G ( t , x ) dt \,. 
\end{equation}
We note that $ T^*  f (s,  x ) := U ( s) f ( x ) $, 
and that the mapping property \eqref{eq:tts} is equivalent to 
\[  \langle T^* T G , F \rangle_{ L^2 ( \RR \times X ) } \leq C 
 \| G \|_{  L^{2} ( \RR , L^1 ( X ) )} \| F \|_{  L^{2} ( \RR , L^1
( X ) ) } \,,\]
which is the same as
\begin{equation}
\label{eq:utus0}
\begin{split} & \left| \int_\RR  \!  \int_\RR  \langle U ( t )^* G ( t ) , U ( s )^* F ( s ) \rangle
\; dt ds \right| 
 \leq C \| G \|_{  L^{2} ( \RR , L^1 ( X ) ) }
\| F \|_{  L^{2} ( \RR , L^1 ( X ) )} \,.
\end{split}
\end{equation}

The hypothesis (with $ \sigma = 1 $) can be restated as 
\begin{equation*}
\begin{split} &   | \langle U ( t )^* G ( t ) , U ( s )^* F ( s )  \rangle |
\leq 
C h^{-1} (h+ | t - s |)^{ -1   } \| G ( t) \|_{ L^{1} ( X ) } 
\| F ( s ) \|_{ L^{1} ( X ) } \,. \end{split} \end{equation*}

Now now apply the Young inequality in $ t $ 
(see \eqref{eq:Young} above) with 
$ p = q=2 $ and $ r = 1 $, noting that 
$ \| \psi ( t ) ( h + |\bullet|)^{-1} \|_{L^1 ( 
\RR )} \leq C \log ( 1/h ) $. That gives \eqref{eq:utus0} completing 
the proof. 

\stopthm

We also need the semiclassical parametrix construction
which is classical [sic!] and where we follow \cite[Appendix a]{HeSj}
-- see also \cite[Proposition 7.3]{SjZw02}, and 
for a textbook presentation \cite[Sect.10.2]{EZ}.
As emphasized in \cite{KoT1} for the dispersive estimates  
of the type used here, we 
only need very basic information about the amplitude, far from 
the precise results needed, for instance, in the study of 
trace formul{\ae} \cite{SjZw02}.

\begin{prop}
\label{p:osc}
Suppose that $ F(t,r)  $ is defined by 
\[  hD_t F( t, r) + P( t) F ( t, r )  = 0 \,, \ \ 
F(r, r) = G(r) ( x, h D) \,, \
G( r )  \in \CIc  ( T^* \RR^k  ) \,. \]
Let us also assume that $ p_t =\sigma ( P ( t) ) $, the Weyl symbol 
(with a possible dependence on $ h$ in the subprincipal symbol part)
of $ P ( t) $, is real.
Then there exists $ t_0 > 0 $, independent of $ h $, such that 
for $ 0 \leq t \leq t_0 $,
\begin{equation}
\label{eqref:13}
F(t,r) u(x)={\frac1 {(2\pi h)^k}}\iint e^{{\frac i  h}(\phi
(t,r,x,\eta )-y\cdot \eta )}b(t,x,\eta ;h)u(y)dyd\eta + E ( t , r ) u ( x ) 
\,, \end{equation}
where
\begin{equation}
\label{eq:14}
\partial _t\phi (t,r,x,\eta )+p_t(x,\partial _x\phi ( t ,r, x, \eta 
)=0,\ \ \ 
\phi (r,r,x,\eta )=x\cdot \eta \,, \end{equation}
$ b \in \CIc ( \RR \times T^* \RR^n ) $, and $ E ( t, r ) =
{\mathcal O} ( h^\infty )  : {\mathcal S}' \rightarrow {\mathcal S} $.
\end{prop}
\begin{proof}
The equation \eqref{eq:14} is the standard eikonal equation 
for which we find a (possibly $h$-dependent) solution $ \phi$. 
The amplitude $ b $  has to satisfy
\[ (hD_t+ p_t^w(x,hD))(e^{i\phi (t,x,\eta )/h}b(t,x,\eta ;h)) =0\,, \]
which is the same as 
\[ (\partial _t\phi +hD_t+e^{-i\phi /h}p_t^w(x,hD)e^{i\phi/h})(b)=0 \,.\]
The Weyl symbol of $e^{-i\phi /h}p_t^w e^{i\phi /h}$
is 
$$q_t(x,\xi )=p_t(x,\phi _x'+\xi )+{\mathcal O}(h^2) \,,$$ 
and using that $\partial _t\phi =-p_t(x,\partial _x\phi )$,
we get
$$(hD_t+  f_t^w(x,h D ) )b={\mathcal O}(h^2) \,,$$ 
with  $f_t(x,\xi )=p_t(x,\phi _x'+\xi )-p_t(x,\phi '_x)$,
and with $\eta $ considered as a parameter. This can be solved 
asymptotically in $h $.
\end{proof}

\begin{prop}
\label{th:sestes}
Suppose that 
$ \chi \in \CIc ( T^* \RR^k ) $, and 
that \eqref{eq:nondeg} holds in $ \supp ( \chi ) $. 
With $ P = p (x, h D) $,  let $ U ( t  ) $
be given by Proposition \ref{p:osc}.
Then for $ \psi \in \CIc ( \RR ) $ with support sufficiently 
close to $ 0 $, and 
\[  U ( t , r ) := \psi ( t ) F ( t , r ) \chi^w ( x , h D ) \ 
\text{ or } \   U ( t , r) := \psi ( t  )  \chi^w ( x, h D ) F ( t , r) \]
we have 
\begin{gather}
\label{eq:sestes}
\begin{gathered}
\sup_{ r \in I } 
\left( \int_{\RR} \|U ( t,r ) f \|_{L^q ( \RR^n ) }^p dt \right)^{\frac{1}p}
\leq B h^{ -\frac 1 p  } \| f \|_{L^2 ( \RR^n ) } \,, \\
 \frac{2}{p} + \frac{ k }{q } = \frac{k}2 \,, \ \ 2 \leq p \leq \infty \,, \ \ 
1 \leq q \leq \infty \,, \ ( p , q ) \neq ( 2, \infty ) \,.
\end{gathered}
\end{gather}
When $ (p , q ) = ( 2, \infty ) $, that is for $ k = 2 $, we have 
the same estimate with $ h^{-1/2 } $ replaced by $ ( \log ( 1/h)/h)^{1/2} $.
\end{prop}
\begin{proof}
In view of Proposition \ref{t:KT} we need to show that  
\begin{equation}
\label{eq:lilt}
\| U ( t , r)  U ( s , r )^* f \|_{L^\infty ( X , dm ) } \leq 
A h^{-k/2} ( h + | t - s |)^{-k/2} \,, \ \ t , s \in \RR \,,
\end{equation}
with constants independent of $ r \in I $. We can put 
$ r = 0 $ in the argument and drop the dependence on $ r $ in 
$ U $ and  $ F $.

We use Proposition \ref{p:osc}
The construction there and the assumption that $ \chi \in \CIc $
show that 
\[ U ( t) =  \widetilde U ( t)  + E ( t ) \,, \]
where 
$$ E ( t) = O ( h^\infty ) : {\mathcal S}' 
\rightarrow {\mathcal S}  \,, $$ 
and the Schwartz kernel of $ \widetilde U ( t) $ is 
\begin{gather}
\label{eq:wideu}
\begin{gathered}
 \widetilde U ( t , x , y ) = 
\frac{1}{(2\pi h)^k} \int_{\RR^k}   e^{{\frac i h}
(\varphi (t,x,\eta )- \langle y, \eta \rangle)} \tilde b(t,y,x,\eta ;h) d\eta 
\,,
\\ \tilde b \in S (1) \cap \CIc ( \RR^{1+3k} ) \,, 
\ \    \varphi ( 0 , x , \eta ) = \langle x , \eta \rangle \,, 
\\ 
\partial_t \varphi ( t,  x , \eta ) + 
p ( t, x ,\partial_x \varphi( t ,x, \eta) ) = 0  \,.
\end{gathered}
\end{gather}

 Hence we only need to prove \eqref{eq:lilt} with $ U $ replaced by 
$ \widetilde U $ and that means that we need an $ L^\infty $ bound
on the Schwartz kernel of $ W ( t, s ) := 
\widetilde U ( t ) \widetilde U ( s ) ^*$:
\[ W( t , s,  x, y ) = 
{\frac{1}{(2\pi h)^{2k}}}\int_{\RR^{3k}}  e^{{\frac i h}
(\varphi (t,x,\eta )- \varphi ( s , y , \zeta ) - \langle z, \eta - \zeta\rangle
)} \; B \; 
d z d\zeta d \eta \,, \]
where
$$ B = B (t,s, x , y,  z ,\eta , \zeta ;h) \in S \cap \CIc ( \RR^{2+6k} ) \,. $$

 The phase is nondegenerate in $ ( z , \zeta ) $ variables 
and stationary for $ \zeta = \eta $, $ z = \partial_\zeta \varphi ( s, y , \zeta ) 
$. Hence we can 
apply the method of stationary phase to obtain
\[ 
 W( t , s , x, y ) = 
{\frac{1}{(2\pi h)^{k}}}\int_{\RR^{k}}  e^{{\frac i h}
(\varphi (t,x,\eta )- \varphi ( s , y , \eta ) )} \; B_1 ( t, s , x, y , \eta ; h ) \; 
d \eta \,, \]
where $ B_1 \in S \cap \CIc ( \RR^{2+3k} ) $.
We now rewrite the phase as follows:
\begin{gather*}  \widetilde \varphi :=  \varphi (t,x,\eta )- \varphi ( s , y , \eta ) = 
( t - s ) p ( 0 , x , \eta )  \\ 
\ \ \ \ +  \langle x - y , \eta + s F (s,  x , y , \eta ) \rangle  +
O ( t - s )^2 \,, \ \ F \in \CI ( \RR^{1+3k} )\,, 
\end{gather*}
where using \eqref{eq:wideu} we wrote 
\[\varphi (s ,x,\eta )- \varphi ( s , y , \eta ) = 
\langle x - y , \eta \rangle + \langle x -y , s F ( s, x, y , \eta) 
\rangle \,. \]

 The phase is stationary when
\[ \partial_\eta \widetilde \varphi = 
( I + s \partial_\eta F ) ( x - y ) + ( t -s ) ( \partial_\eta p + 
O ( t - s) ) = 0 \,, \]
and in particular, for $ s $ small, having a stationary point implies
\[  x - y = O ( t - s ) \,, \]
as then $  ( I + s \partial_\eta F ) $ is invertible. The Hessian is given by 
\[ \begin{split}
 \partial^2_\eta \widetilde \varphi & = s \partial^2_\eta F ( x - y ) 
+ ( t -s ) ( \partial^2_\eta p + O ( t - s ) ) \\
& =  ( t - s ) ( \partial^2_\eta p + O ( |t|  + |s| ) )  
\,, \end{split} \]
where $ \partial^2_\eta p = \partial^2_\eta p ( 0 , x , \eta) $. 

 Hence, for $ t $ and $ s $ sufficiently small, that is for a
suitable choice of the support of $ \psi $ in the definition of 
$ U ( \bullet ) $, the nondegeneracy assumption \eqref{eq:nondeg}
implies that at the critical point
\[  \partial_\eta^2 \widetilde \varphi =  ( t - s ) \psi ( x , y )  \,.
\]
Hence for $ | t - s | > M h $ for a large constant $ M $ we can 
use 
the stationary phase estimate to obtain
\[ | W ( t , s,  x, y ) | \leq  C h^{-k/2} ( h + | t -s |) ^{-k/2} \,.\]
When $ | t - s | < M h $ we see that the trivial estimate of the 
integral gives
\[ | W ( t , s,  x, y ) | \leq C h^{-k} \leq C'   h^{-k/2}
 ( h + | t -s |)^{-k/2} 
\,, \]
which is what we need to apply Proposition \ref{t:KT}.
\end{proof}

\section{$L^p$ estimates in the nondegenerate principal type case}
\label{Lpn}

In this section we prove the general version of the
part of Theorem \ref{th:1}, in which $ V ( x ) \neq 0 $. 
That covers the case of spectral problems on 
 Riemannian manifolds in which case we take $ V ( x ) \equiv -1 $. 

To state the general result 
we formulate the following nondegeneracy assumptions 
at $ ( x_0 , \xi_0 ) \in T^* \RR^n $:
\begin{equation}
\label{eq:non1}
 p ( x_0 , \xi_0 ) = 0 \; \Longrightarrow \; 
 \partial_\xi p ( x_0, \xi_0 ) \neq 0 \,. 
\end{equation}
Then the set 
\[ {\rm Char}_{x_0} ( p ) \defeq \{ \xi \; : \;  p ( x_0 , \xi) =0 \}\,, 
\] 
is a smooth hypersurface in $\RR^n $. We then assume that 
\begin{equation}
\label{eq:microhyp}
 \text{the second fundamental form of 
$ {\rm Char}_{x_0} ( p ) $ is nondegenerate at $ \xi_0 $. } 
\end{equation}

In more concrete terms, by a linear change of variables,  we can assume that $ 
\partial_\xi p ( x_0, \xi_0) = ( \rho , 0, \cdots , 0 ) $, $ \rho \neq 0$. Then
near $ (x_0,\xi_0) $, 
\begin{equation}
\label{eq:mss}  p ( x , \xi ) = e( x , \xi) ( \xi_1 - a ( x, \xi') ) \,,  \ \ e ( x_0, 
\xi_0 ) \neq 0 \,, \end{equation}
and our assumption is
\begin{gather}
\label{eq:microhyp1}
\begin{gathered}
\partial^2_{\xi'} a ( x_0 ,\xi'_0 )  \ \text{ is nondegenerate. }
\end{gathered}
\end{gather}
As in the remark following \eqref{eq:nondeg} we note that this 
assumption is invariant under linear changes of coordinates in 
$ \xi $.
In particular \eqref{eq:microhyp1}
is invariant under changes of variables. We should mention here
that symbol factorizations \eqref{eq:mss} have a long tradition 
in microlocal analysis
and in the context of $ L^p$ estimates were used in \cite{MSS}.

\begin{thm}
\label{th:seso}
Suppose that $ u ( h )  $, $ \| u ( h ) \|_{ L^2 } = 1 $, 
is a family of functions
satisfying the frequency localization condition \eqref{eq:loc}.
Suppose also that \eqref{eq:non1} and 
\eqref{eq:microhyp} are satisfied on $ \supp \chi $.

Then for $ p = 2 ( n + 1 ) / ( n- 1) $, and any $ K \Subset \RR^n $,
\begin{equation}
\label{eq:sogge1}
\| u ( h ) \|_{ L^p  } \leq  Ch^{-1/p} \left( \| u ( h ) \|_{2} + 
\frac 1 h \| p ( x , h D )  \|_{ L^{2} } \right)
\,.
\end{equation}
\end{thm}

\bigskip

\noindent
{\bf Remark.} The first example in the remark after Theorem \ref{th:8.5}
shows that the curvature condition \eqref{eq:microhyp} is
in general necessary. In fact, if $ p( x, hD) = h D_{x_1} $ and 
$$ u ( h ) = h^{-(n-1)/2} \chi ( x_1 ) \chi ( x'/h ) $$
 then 
for $ p = 2 ( n + 1) /( n-1) $, 
\[ \| u \|_{ L^p } \simeq h^{ (n-1)(1/p-1/2) }
= h^{-(n-1)/(n+1)} \neq  O ( h^{-1/p } ) 
\,. \]
However for the simplest case in which \eqref{eq:microhyp} holds,
\[ p ( x, \xi ) = \xi_1 - \xi_2^2 - \cdots - \xi_n^2 \,, \]
the estimate \eqref{eq:sogge1} is optimal. To see that put
\[ u ( h ) := h^{-(n-1)/4} \chi_0 (x_1 ) \exp ( - |x'|^2/2h)  \,, \]
where $ x = (x_1 , x') $, $ \chi_0 \in \CIc ( \RR )$.
Then 
$$ ( - h^2 \Delta_{x'} + |x'|^2 ) u ( h ) = (n-1) h \; u ( h ) \,, $$
$ \| u ( h ) \|_{2}  \simeq 1$,  $ |x'|^{2k} u ( h ) = O_{L^2} ( h^k ) $.
Hence,
\[ p^w ( x ,h D ) u ( h ) = O_{L^2} ( h ) \,, \]
and 
\[ \| u ( h ) \|_{p ( \RR^n ) } \simeq h^{ (n-1)(2/p - 1 )/4 } 
= h^{-1/p } \,, \ \ p = 2(n+1)/(n-1) \,.\]

Before proving Theorem \ref{th:seso} we prove a lemma which is a
consequence of Proposition \ref{th:sestes}:

\begin{lem}
\label{l:ff}
In the notation of Proposition \ref{th:sestes} and for 
\[ p = q = \frac{ 2 ( k + 2 ) }{ k } \,, \]
we have
\begin{equation}
\label{eq:ff}
 \| \int_{0}^t U ( t , s ) {\bf 1}_{ I} ( s)
f ( s , x) ds \|_{ L^p ( \RR_t \times \RR^{k}_x )} \leq C h^{-1/p} \int_{\RR } 
\| f ( s, x ) \|_{ L^2 ( \RR^{k}_x ) } d s \,.
\end{equation}
\end{lem}
\begin{proof}
 We apply the integral version of Minkowski's
inequality and \eqref{eq:sestes}:
\[ \begin{split} & \| \int_{0}^t U ( t , s ) {\bf 1}_{ I} ( s)
f ( s , x) ds \|_{ L^p ( \RR_t \times \RR^{k}_x )} \\
& \ \ \ \ \ \leq C
\int_{ I \cap \RR_+ }  \|  {\bf 1}_{ [s, \infty)} ( t)  U ( t , s )
f ( s , x ) \|_{ L^p ( \RR_t \times \RR^{k}_x ) } ds \\
&  \ \ \ \ \ \leq C
\int_{ I \cap \RR_+ }  \| U ( t , s ) f ( s , x )
\|_{ L^p ( \RR_t \times \RR^{k}_x ) } ds \\
&  \ \ \ \ \  \leq C' h^{-1/p} 
\int_I \| f ( s, x )\|_{ L^2 ( \RR^k_x ) } ds \,.
\end{split} \]
\end{proof}

\noindent
{\em Proof of Theorem \ref{th:seso}:} We follow the same procedure 
as in the proof of Theorem \ref{th:8.5} but replacing the energy 
estimate of Lemma \ref{l:simple} with the Strichartz estimate.

We factorize $ p ( x , \xi ) $ as in \eqref{eq:pel} and we easily 
conclude that for $ \chi $ with sufficiently small support,
\[   ( h D_{x_1} - a ( x , hD_{x'} ) ) ( \chi^w u ( h ) ) = O_{L^2} ( h ) 
\,. \]
Let 
\[ f ( x_1 , x', h ) =    ( h D_{x_1} - a ( x , hD_{x'} ) )( \chi^w u ( h ) ) 
\,. \] 
Since $ \| f \|_{2} = O(h) $, we see 
\begin{equation}
\label{eq:ff1} \int_\RR \| f ( x_1 , \bullet) \|_{L^2 ( \RR^{n-1} ) } dt \leq 
C \| f \|_{ L^2 ( \RR^n ) } = O ( h ) \,. \end{equation}

 We now apply Proposition  \ref{th:sestes} with $ t = x_1 $ and 
$ x $ replaced by $ x' \in \RR^{n-1} $, that is $ k = n-1$.
We also take
 $ p = q $ in \eqref{eq:sestes},
\[ p = q = \frac{ 2 ( k + 2) }{ k } = \frac{ 2 ( n+1)}{n-1} \,.\] 
The assumption \eqref{eq:microhyp} shows that $ \partial_{\xi'}^2 a 
$ is nondegenerate in the support of $ \chi $. We can choose $ \psi $
and $ \chi $ in the definition of $ U ( t , s ) $ in the statement of
Proposition \ref{th:sestes} so that 
\[  \chi^w ( x , h D ) u ( x_1, x', h ) = \frac{i}{h} \int_{0}^{x_1} 
U ( x_1,s) f ( s,
x' ) ds + O_{{\mathcal S}} (h^\infty ) \,.\]

Then, using Lemma \ref{l:ff},
\[ \begin{split}
\| \chi^w ( x , h D ) u \|_{ L^p } & \leq 
\frac{1}{h} \, h^{-1/p} \, \int_\RR \| f ( s, \bullet , h ) \|_{ L^2 ( \RR^{n-1} ) }ds  + O ( h^\infty )  \\ & = O ( h^{-1/p} ) \,.\end{split}
\]
A partition of unity argument used in the proof of Theorem \ref{th:8.5}
concludes the proof.
\stopthm

\section{$L^p$ estimates in the nondegenerate non-principal type case}
\label{s:nonpr}

In this section we prove the general result corresponding to 
the part of Theorem \ref{th:1} giving estimates near
points where $ V ( x ) = 0 $. 
This means considering the case of $ d_\xi p (x_0 , \xi_0 ) = 0$. 
For functions localized near $ (x_0 , \xi_0 ) $ 
in the sense of \eqref{eq:loc}, the estimates will hold under 
the following nondegeneracy condition at $ ( x_0 , \xi_0 ) $:
\begin{equation}
\label{eq:nondeg}
\partial_\xi^2 p ( x_0 , \xi_0 ) \ \text{ is non-degenerate .}
\end{equation}
We then have
\begin{thm}
\label{th:nonpr}
Let $ n > 2 $, suppose that the localization condition \eqref{eq:loc} holds
and that $ \supp \chi $ is a small neighbourhood of a point $ ( x_0 , \xi_0 ) $
at which $ p ( x_0, \xi_0 ) = 0 $, $ d_\xi p ( x_0 , \xi_0 ) = 0 $, 
and \eqref{eq:nondeg} holds.
Then 
\begin{equation}
\label{eq:4} \| u ( h ) \|_{ q} \leq C h^{-1/2} \left( 
\| u ( h ) \|_{2} + \frac 1 h \| p ( x , h D ) u ( h ) \|_{ 2}  \right) \,, \ \ q = \frac{ 2 n }{ n-2} \,.
\end{equation}
Also,
\begin{equation}
\label{eq:44} \| u \|_{\infty } \leq h^{-(n-1)/2 } 
 \left( 
\| u ( h ) \|_{2} + \frac 1 h \| p ( x , h D ) u ( h ) \|_{ 2} \right) \,. 
\end{equation}
When $ n = 2 $ the same estimate holds with $ h^{-1/2} $ 
replaced by $ (\log ( 1/h )/h)^{1/2} $.
\end{thm}
\begin{proof}
To simplify the proof we assume that \eqref{eq:nondeg} holds
on the support of $ \chi $, in other words,
\begin{equation} 
\label{eq:3} ( x , \xi ) \in \supp \chi 
\Longrightarrow \ 
\det \partial_\xi^2 p ( x , \xi ) \neq 0 \,. 
\end{equation} 
The Hessian, $ \partial_\xi^2 f ( \xi_0 ) $, of a 
smooth function $ f ( \xi ) $ is not invariantly defined unless
$ \partial_\xi f ( \xi_0) = 0 $. 
However the statement \eqref{eq:nondeg} is 
invariant if only {\em linear} transformations in $ \xi $ are 
allowed. That is the case for symbol transformation 
induced by changes of variables in 
$ x $, see \cite[Theorem 8.1]{EZ}.

Suppose that $ P u = h f $ and that the assumptions
of theorem hold. In particular, $ f \in L^2 $ and $ 
\chi ( x , h D ) f = f + {\mathcal O}_{\mathcal S} ( h^\infty ) $. 
Then 
\[ ( hD_t + P ) u = h f \,. \]
Using the notation of Proposition \ref{p:osc}, 
Duhamel's formula gives
\[ \psi ( t ) u ( x ) = 
U ( t ,0) u ( x ) + i \int_0^t U ( t , s ) f ( x ) ds +
{\mathcal O}_{\mathcal S} ( h^\infty )\,. \]
Choose $ I \Subset \RR $ so that $ \supp \psi \subset I $.
Propositions \ref{th:sestes}
applied with $ p = 2 $ and $ q = 2n/(n-2) $, and the integral 
version of Minkowski's inequality, show that 
\begin{equation}
\label{eq:qqq} 
\begin{split} \left( \int_{\RR} \psi( t )^2 dt \right)^{\frac12} 
\| u \|_{q} & \leq C h^{-\frac12}  \| u \|_2 + C \left( \int_I \left \| 
\int_0^t U ( t , s ) f ( x) ds \right\|_{ q}^2 dt  \right)^{\frac12} 
\\
& \leq 
 C h^{-\frac12}  \| u \|_2 + C \left( \int_I  
\int_0^t \| U ( t , s ) f ( x)  \|_{ q}^2 ds dt  \right)^{\frac12} \\
& \leq 
 C' h^{-\frac12}(   \| u \|_2 + \| f \|_2 ) \,. 
\end{split}
\end{equation}
This proves \eqref{eq:4}. To see \eqref{eq:44} we use \eqref{eq:4}, the
localization assumption \eqref{eq:loc}, and Lemma \ref{l:3}:
$ \| u \|_{\infty } \leq h^{ -n/q } \| u \|_q $, $ 
n/q + 1/2 = n/(2n/(n-2)) + 1/2 = (n-1)/2 $.

For $ n = 2 $ we use the weaker version of the end point 
result in Proposition \ref{th:sestes}.
\end{proof}

\medskip
\noindent
{\bf Remark.} We should stress that to obtain 
\eqref{eq:44} we do not need the subtle end point Strichartz estimate
but its easier interior version: the same proof 
based on that gives
\[ 
 \| u ( h ) \|_{ q} \leq C h^{-(n-1)/2+n/q} \left( 
\| u ( h ) \|_{2} + \frac 1 h \| p ( x , h D ) u ( h ) \|_{ L^{2} 
} \right) \,, \ \  \frac{ 2 n }{ n-2} < q < \infty \,,
\]
from which the $L^\infty $ estimate follows in the same way.

\medskip

In the generality we work in the bound \eqref{eq:44} 
is {\em not} true for $ n = 2$.
Consider the following operator
\begin{equation}
\label{eq:count}
  P = p ( x , h D ) \,, \ \ p ( x , \xi ) = \xi_1^2 - \xi_2^2 + 
x_1^2 - x_2^2 \,. 
\end{equation}

Let $ w_\ell ( x ) $ be the normalized eigenfuction of $ D_y^2 + y^2 $ in 
dimension one with eigenvalue $ 2 \ell + 1$. Then for $ \ell = 2j $ even
we have the classical fact based on Stirling's approximation:
\[ \begin{split} | w_{2j} ( 0 ) | & = 
\frac{ 1 \cdot 3 \cdots ( 2j - 3 ) \cdot ( 2j - 1 ) }
{ \sqrt{ ( 2j)!} \pi^{\frac14} } 
= \frac{ \sqrt { ( 2j)!} }{ 2^j j! \pi^{\frac14} } \simeq 
\frac{ ( ( 2j)^{2j+\frac12} e^{-2j} )^{\frac12} }{ 2^j j^{j+\frac12} 
e^{-j} }
= j^{-\frac 1 4 } 
\,, \end{split} \]
and we can choose 
$ w_{2j} $ to be real and to satisfy $ w_{2j} ( 0 ) > 0 $.
We now put
\[  v_k = \frac 1 {\sqrt k } \sum_{\ell=1}^k \left( 2^{- \ell /
2  } \sum_{ j=2^{\ell} }^{ 2^{\ell+1} - 1} w_{2j} ( x_1 ) w_{2j} (x_2 ) \right)\,.\]
Since all the different summands are 
orthogonal we have $ \|v_k \|_{2}  = 1 $, and
\[ \begin{split} \| v_k \|_{\infty} & \geq 
 \frac 1 {\sqrt k } \sum_{\ell=1}^k \left( 2^{- \ell /
2  } \sum_{ j=2^{\ell} }^{ 2^{\ell+1} - 1} u_{2j} ( 0 ) u_{2j} (0) \right) 
\simeq
 \frac 1 {\sqrt k } \sum_{\ell=1}^k \left( 2^{- \ell /
2  } \sum_{ j=2^{\ell} }^{ 2^{\ell+1} - 1} j^{-1/2}  \right) 
\simeq k^{\frac12} \,. 
\end{split} 
\]
Now put 
\[  u ( h ) = h^{ - \frac 12} v_k ( x/ h^{\frac12} ) \,, \ \ 
 2^{ -k } \leq h \leq 2^{-k+1} \,.\]
With $ P $ given by \eqref{eq:count}, $  P u ( h ) =  0 $, and
\[  \| u ( h ) \|_\infty \geq ( \log ( 1/ h ) / h ) ^{ \frac 12} 
= \left( h^{ - \frac 12} \| u \|_{2} \right) \log ( 1/h )^{\frac12} \,. \]
Since we have 
\[ \psi ( (h D)^2 + x^2 ) ) u = u \,, \]
with $ \psi \in \CIc $, the localization condition in Theorem \ref{th:nonpr}
follows.

\section{Improved estimates for Schr\"odinger operators.}
\label{isch}

In this section we prove a reformulation 
of Theorem \ref{th:new}:

\begin{thm}
\label{th:newg}
Let $ p ( x, \xi ) $ be of the form
\[ p ( x , \xi ) = \sum_{ i , j =1}^n a_{ij} ( x ) \xi_i \xi_j 
+ V ( x ) \,.\]
Suppose that $ u ( h ) $ satisfies the localization condition 
\eqref{eq:loc} and that $ \supp \chi $ is a small neighbourhood
of $ ( x_0, \xi_0 ) $ at which
\begin{equation}
\label{eq:13} p ( x_0, \xi_0 ) = 0 \,, \ \ d_\xi p ( x_0, \xi_0 ) = 0 \,, \ \
d_x p ( x_0, \xi_0 ) \neq 0 \,, \ \ \partial_\xi^2 p ( x_0 , \xi_0 ) 
\ \text{is positive definite.} \end{equation}
Then 
\[   \| u \|_2 + \frac1 h \| p ( x , h D ) u \|_2 = {\mathcal O} ( 1 ) \]
implies estimates \eqref{eq:thnew}.
\end{thm}

\noindent
{\bf Remark.} It seems clear that the assumption \eqref{eq:13}
is sufficient for the conclusion of the theorem to hold. We
restrict ourselves to the special case of quadratic 
hamiltonians in order to streamline the rather involved proof. On the other
hand the case of a nondegenerate but not necessarily 
definite Hessian $ \partial^2_\xi p $ poses a greater
challenge. 

We start with a reduction of the problem. 
We can assume that $ (x_0, \xi_0 ) = ( 0 , 0 ) $, and since we 
work locally, and $ d V( 0 ) \neq 0 $, we can change coordinates
so that $ V ( x) = - x_1 $. We can then choose normal geodesic
coordinates for the quadratic form $ g ( x, \xi ) = \sum_{ i,j}
a_{ij} ( x ) \xi_i \xi_j $ with respect to the surface $ x_1 = 0 $.
That means that we can replace $ p $ with 
\begin{gather}
\label{eq:formp} 
\begin{gathered}
p ( x , \xi ) = \xi_1^2 + \lambda ( x , \xi') - c ( x ) x_1 \,, \\ 
\lambda ( x , \xi' ) = \sum_{ i,j=2}^{n} \tilde a_{ ij} ( x ) \xi_i 
\xi_j \geq \frac{1}{C} |\xi'|^2 \,, \ \ c ( 0 ) = 1 \,. 
\end{gathered}
\end{gather}
The Hamilton vector field of $ p $ is 
\begin{equation}
\label{eq:Hp}
 H_p = 2 \xi_1 \partial_{x_1} + ( c ( x ) - x_1 \partial_{x_1} c ( x) 
- \partial_{x_1} \lambda ( x , \xi') ) \partial_{ \xi_1 } + V \,,
\end{equation}
where the vector field $ V $ does not involve differentiation with 
respect to $ x_1 $ and $ \xi_1 $. At $ ( x, \xi ) = ( 0 , 0 ) $
we have $ H_p - V = 2 \xi_1 \partial_{x_1} + \partial_{\xi_1 } $
and this model vector field is essential in the argument.

We observe that for $ x_1 < - \delta < 0 $ the operator is 
elliptic in the semiclassical sense, while for $ x_1 > \delta > 0 $
we can apply Theorem \ref{th:seso} which gives a stronger conclusion 
than \eqref{eq:thnew}. The analysis is confined to a small 
neighbourhood of $ x_1 = 0 $ and we will obtained estimates in 
regions defined by 
$ -1 < x_1 < \epsilon $. On the energy surface, $ p = 0 $,
this implies that $ | \xi | < C \epsilon^{1/2} $ and the 
uncertainty principle gives a natural restriction on $ \epsilon $:
$ \epsilon \times \epsilon^{1/2} \leq K h $, that is, 
$ \epsilon \leq M h^{2/3} $. 

We start with the following
\begin{lem}
\label{l:7.1}
Let $ P = p^w ( x, h D ) $ with $ p $ given by \eqref{eq:formp},
and suppose that $ u $ is supported in a small neighbourhood of 
$ 0 $. Define 
\[ \Omega_\epsilon = \{ ( x_1 , x' ) \; : \; x_1 < \epsilon \} \,.\]
Then, for $ \epsilon > h^{2/3} $, 
\begin{equation*}
\| ( h D)^\alpha u \|_{ L^2 ( \Omega_\epsilon ) } 
\leq C \epsilon^{\frac12}  
\| u \|_{ L^2 ( \Omega_{2 \epsilon } )  } + 
 C \epsilon^{-\frac12}   
\| P u \|_{ L^2 ( \Omega_{2 \epsilon } )  } 
\,, \ \ | \alpha| = 1 \,.
\end{equation*}
\end{lem}
\begin{proof}
Let us put $ u_\epsilon = \chi ( x_1/\epsilon ) u $ where 
$ \chi \in \CI ( \RR , [ 0 , 1] ) $ is supported in $ t < 2 $ and
is equal to $ 1 $ in $ t \leq 1 $. Then 
\[ P u_ \epsilon = \chi ( x_1/\epsilon ) P u + 
\frac{2}{i} \frac{h} \epsilon \chi' ( x_1/ \epsilon ) h D_{x_1} u 
- \frac{ h^2 } {\epsilon^2 } \chi'' ( x_1 / \epsilon ) u \,. \]
Integration by parts gives
\[  \langle \chi'( x_1 /\epsilon ) h D_{x_1} 
u , u_\epsilon \rangle  = {\mathcal O} ( h/ \epsilon ) \| u \|^2_{ L^2 ( 
\Omega_{2\epsilon } )} \,, \]
and hence 
\[ \begin{split}
 \langle P u _\epsilon , u_\epsilon \rangle /
\epsilon & = 
{\mathcal O} ( 1/\epsilon )  \| u \|_{ L^2 ( \Omega_{2 \epsilon } )  }
\|P u \|_{ L^2 ( \Omega_{2 \epsilon } )  }  
+ {\mathcal O} ( h^2/ \epsilon^3 )  \| u \|_{ L^2 
( \Omega_{2 \epsilon } )  }^2  \\ 
& = 
{\mathcal O} ( 1/\epsilon )  \| u \|_{ L^2 ( \Omega_{2 \epsilon } )  }
\|P u \|_{ L^2 ( \Omega_{2 \epsilon } )  }  
+ {\mathcal O} ( 1)  \| u \|_{ L^2 
( \Omega_{2 \epsilon } )  }^2   \\
& =  
{\mathcal O} ( 1/\epsilon^2 )  
\|P u \|^2_{ L^2 ( \Omega_{2 \epsilon } )  }  
+ {\mathcal O} ( 1)  \| u \|_{ L^2 
( \Omega_{2 \epsilon } )  }^2 
\,,
\end{split} \]
where we used $ h^2 / \epsilon^3 \leq 1 $.

On the other hand \eqref{eq:formp} shows that for any $ |\alpha|=1$, 
\[ \begin{split}
\frac{1} \epsilon \langle P u_\epsilon , u_\epsilon \rangle
& \geq \frac{1}{C \epsilon } \| ( h D )^\alpha u_\epsilon \|^2 - 
C \langle ( x_1 /\epsilon ) u_\epsilon , u_\epsilon 
\rangle
 \geq \frac{1}{C \epsilon } \| ( h D )^\alpha u_\epsilon \|^2 - 
C  \| u \|_{ L^2 ( \Omega_{2 \epsilon } )  }^2 \,.
\end{split} \]
Thus,
\[ \sum_{ |\alpha| = 1 } \| ( h D )^\alpha u_\epsilon \|^2 
\leq C  \epsilon  
\| u \|_{ L^2 ( \Omega_{2 \epsilon } )  }^2 
+ \frac C \epsilon \| P u \|_{ L^2 ( \Omega_{2 \epsilon } )  }^2 
\,.\]
which proves the lemma.
\end{proof}
We remark that a similar integration by parts argument
gives a global weighted estimate (see \cite[(13)]{KoT} for
a slightly weaker version in a particular case):
\begin{equation}
\label{eq:globw}
\| ( x_1^2 + \epsilon^2 )^{-\frac14} h D u \| \leq 
C \| u \| + C \epsilon^{-1} \| P u \| \,, \ \ \epsilon > 0 \,, \ \ 
0 < h < 1 \,.
\end{equation}
In fact, 
\[ \begin{split}
 \| u \|^2 +  \epsilon^{-2} \| P u \|^2  & \geq  
\langle ( x_1^2 + \epsilon^2 )^{-\frac12} u , P u \rangle 
\\ &  = \| ( x_1^2 + \epsilon^2)^{-\frac14} h D_{x_1} u \|^2 + 
\langle  ( x_1^2 + \epsilon^2)^{-\frac12}\lambda^w ( x , h D_{x'}) u , u 
\rangle \\ 
& \ \ \ \ \ \ \ 
+ \langle [ h D_{x_1 } , ( x_1^2 + \epsilon^2)^{-1/2} ] u , 
h D_{x_1} u \rangle - \langle x_1 ( x_1^2 
+ \epsilon^2 )^{-\frac12} u , u \rangle \\
 &   \geq  \frac{1}{C}  \sum_{ |\alpha| = 1 } 
\| ( x_1^2 + \epsilon^2)^{-\frac14}  ( h D )^\alpha u \|^2 -  
C \| u\|^2 \,. \end{split} \]
For estimating the commutator term we noticed that
\[ \begin{split}
 | \langle [ h D_{x_1 } , ( x_1^2 + \epsilon^2)^{-1/2} ] u , 
h D_{x_1} u \rangle | & \leq h \| x_1 / ( x_1^2 + \epsilon^2 )^{ } u \|
\| ( x_1^2 + \epsilon^2)^{ -1/4} h D_{x_1} u \| \\
& \leq h ( C \| u \|^2 +  \| ( x_1^2 + \epsilon^2)^{ -1/4} h D_{x_1} u 
\| / C ) \,. \end{split} \]

The next lemma is a preparation for a positive commutator
argument:
\begin{lem}
\label{l:7.2}
In the notation of Lemma \ref{l:7.1}, let 
\begin{equation*}
 A \defeq \frac{1}2 ( \epsilon^{-\frac12} \alpha ( x_1 / \epsilon )
h D_{x_1} + (h D_{x_1} )  \epsilon^{-\frac12} \alpha ( x_1 / \epsilon )
) \,, \ \ \epsilon \geq h^{2/3} \,,
\end{equation*}
where $ \alpha \in \CI ( \RR ) $, 
$ \alpha ( t) = 1 $, for $ t \leq 1 $, and for all $ k\in \NN$,
$ \partial^{k} \alpha ( t ) = {\mathcal O} ( t^{-1/2-k} ) $, $ t > 1 $.

Suppose that $ u $ satisfies \eqref{eq:loc} with $ \chi $ is supported
near $ ( 0 , 0 ) $, $ \| u \| = {\mathcal O} ( 1 ) $. 
and $ \| P u \| = {\mathcal O} ( h ) $. Then
\begin{equation}
\label{eq:l72}
\frac i h \langle [ P , A ] u , u  \rangle = 
 \epsilon^{-\frac12} \left\langle\left( ( 2/{\epsilon} )
h D_{x_1} \alpha' \left( { x_1}/{\epsilon} \right) h D_{x_1}  + 
c ( x ) \alpha \left( { x_1}/{\epsilon} \right)  u 
 \right) , u \right\rangle + {\mathcal O} ( 1 ) \,.
\end{equation}
\end{lem}
\begin{proof}
The operator $ ( i/h ) [ P , A ] $ is a second order selfadjoint
operator and a computation gives 
\[ \frac i h [ P , A ] = 
\epsilon^{-\frac12}\left( h D_{x_1} \alpha' \left( { x_1}/{\epsilon} \right) h D_{x_1}  + 
c( x ) \alpha \left( { x_1}/{\epsilon} \right)  
(1- x_1 \partial_{x_1} c ( x) - \partial_{x_1} \lambda^w 
( x , h D_{x'} ) )
\right) \,,\]
(this also follows from the composition 
formula in Weyl calculus using \eqref{eq:Hp}). We need to show that
for $ u $ satisfying our assumptions we have 
\begin{equation}
\label{eq:lem721}
\langle \epsilon^{-\frac12} \alpha \left( { x_1}/{\epsilon} \right)  
x_1 u , u \rangle = {\mathcal O} (1) \,,
\end{equation}
and 
\begin{equation}
\label{eq:lem722}
\langle \epsilon^{-\frac12} \alpha \left( { x_1}/{\epsilon} \right) 
  \partial_{x_1} \lambda^w 
( x , h D_{x'} ) u , u \rangle = {\mathcal O} ( 1 ) \,.
\end{equation}
To see \eqref{eq:lem721} we note that for $ x_1 \geq  \epsilon
 $ we have 
\[  \epsilon^{\frac12} x_1 \alpha ( x_1/ \epsilon ) = 
x_1^{\frac12} ( x_1 / \epsilon )^{1/2} \alpha ( x_1/ \epsilon )
= {\mathcal O} ( 1 ) \,, \]
since $ \alpha ( t ) = {\mathcal O} ( \langle t \rangle^{-1/2} ) $
for $ t \geq 1  $ . For $ x_1 \leq \epsilon $ we proceed as in the 
proof of Lemma \ref{l:7.1} using the favourable sign of 
$ x_1 $ in the equation: in the notation used there
\[ 
\begin{split}
-  \langle \epsilon^{-\frac12} \alpha \left( { x_1}/{\epsilon} \right)  
x_1 u_\epsilon , u_\epsilon \rangle & = {\mathcal O} ( \epsilon^{\frac12})
\| u_\epsilon \|^2 - 
 \langle \epsilon^{-\frac12} 
( (hD_{x_1})^2 + \lambda^w ( x , h D_{x'} ) ) u_\epsilon , 
u_\epsilon \rangle \\
& \leq C( \epsilon^{\frac12} +  h/\epsilon^{\frac12}  )  
\leq C \epsilon^{\frac12} \,,
\end{split}\]
since, by the sharp G{\aa}rding inequality (note that we are near 
frequency $ 0 $), or by integration by parts, 
\[ ( (hD_{x_1})^2 + \lambda^w ( x , h D_{x'} ) ) v  , 
v \rangle \geq - h \| v\|^2 \,. \]

To see \eqref{eq:lem722} we integrate by parts in the $ x' $ 
variables to obtain 
\[ \left|
\langle \epsilon^{-\frac12} \alpha \left( { x_1}/{\epsilon} \right) 
  \partial_{x_1} \lambda^w 
( x , h D_{x'} ) u , u \rangle \right| \leq 
C \epsilon^{-\frac12} 
\sum_{ | \alpha  | = 1 } 
\| |\alpha \left( { x_1}/{\epsilon} \right) |^{\frac12} (h D)^{\alpha}
 u \|^2 + {\mathcal O} ( h/\epsilon^{\frac12} ) \,.
\] 
The last term came from commutators and we used the fact that
$ \| u \|, \| ( h D )^{\alpha } u \| = {\mathcal O} ( 1)  $ 
(see the assumption \eqref{eq:loc}). 
For $ x_1 < \epsilon  $ we use Lemma \ref{l:7.1} which gives
\[ \|  (h D)^{\alpha} u \|_{ L^2 ( \Omega_\epsilon ) }  \leq 
C \epsilon^{\frac12}  + C h \epsilon^{-\frac12} = {\mathcal O} 
( \epsilon ^{\frac12} )\,. \]
For $ x_1 > \epsilon $ we have 
\[  \epsilon^{-\frac14} 
|\alpha \left( { x_1}/{\epsilon} \right) |^{\frac12} \leq 
C ( x_1^2 + \epsilon^2 )^{-\frac1 8} \leq C ( x_1^2 + \epsilon^2)^{
-\frac 14} \,, \]
and the estimate follows from \eqref{eq:globw}.
\end{proof}

The next lemma is our crucial estimate. Heuristically, as has been 
explained in \cite[Sect.3]{KoT}, it follows from estimating
the length of trajectories on the energy surface over the set
$ x_1 < \epsilon$: that length is at most $ \epsilon^{\frac12}$.
To make this rigorous we apply the standard positive commutator
argument but with an $ \epsilon $ dependent multiplier.

\begin{lem}
\label{l:73}
Under the assumptions on $ u $ and $ \epsilon $ from  Lemma \ref{l:7.2},
we have 
\begin{equation*}
\| ( h D)^{\alpha } u \|_{ L^2 ( \Omega_{ \epsilon } )  }
= {\mathcal O} ( \epsilon^{ 1/4+ | \alpha |/2 } ) \,, \ \ 
| \alpha | \leq 1 \,. 
\end{equation*}
\end{lem}
\begin{proof}
In view of Lemma \eqref{l:7.1} we only need to prove 
the estimate for $ \alpha = 0 $. We will apply 
Lemma \ref{l:7.2} with 
\[  \epsilon \geq M h^{2/3} \,, \ \ M \gg 1 \,,\]
and $ \alpha ( t ) $ such that 
\begin{equation}
\label{eq:propa}
\begin{split}
& 2 t \alpha' ( t ) + \alpha ( t ) \geq 1 \,, \ \text{ for } \ 
 t \leq 1 \,, \\
& 2 t \alpha' ( t ) + \alpha ( t ) \geq ( 1 + |t|)^{-3/2}  \,, \ \ 
\alpha' ( t )\leq 0 \,, \ \text{ for } \   t \in \RR \,. 
\end{split} 
\end{equation}
We construct such a function by smoothing out
\[ \alpha_0 ( t ) \defeq \left\{ \begin{array}{ll} 
\ \ 1 & t \leq 1 \,, \\
2/ \sqrt{ t +1 } & t \geq 1 \,. \end{array} \right. \]
We first observe that Lemma \ref{l:7.1} and the global estimate
\eqref{eq:globw} show that $ A u = {\mathcal O}_{L^2} ( 1 ) $.
In fact, 
\[ A u = \epsilon^{-\frac12} \alpha ( x_1/\epsilon ) h D_{x_1} u 
- i h \epsilon^{-\frac32} \alpha' ( x_1/\epsilon ) u =
{\mathcal O} ( (x_1)_+^2 + \epsilon^2 )^{-\frac14} h D_{x_1} u 
+ {\mathcal O}_{L^2} ( 1 ) \,. \]
We can then use Lemma \ref{l:7.1} in $ x_1 < \epsilon $ and 
the estimate \eqref{eq:globw} for $ x_1 \geq 0 $. 

Since we assumed that $ P u =  {\mathcal O}_{L^2} ( h ) $, as 
$ P $ and $ A $ are selfadjoint,  we have 
\[ \mathcal O ( 1) = \Re \frac{h} i \langle P u , A u \rangle 
= \Re \frac{h } i \langle [ P , A ] u , u \rangle \,.\]
From \eqref{eq:l72} we then have 
\[ \left\langle\left( ( 2/{\epsilon} )
h D_{x_1} \alpha' \left( { x_1}/{\epsilon} \right) h D_{x_1}  + 
c ( x) \alpha \left( { x_1}/{\epsilon} \right)  u  \right) , u \right\rangle
= {\mathcal O} ( \epsilon^{\frac 12 } ) \,, \]
and we want to estimate the left hand side from below. 
For that we rewrite it as 
\begin{equation}
\label{eq:form}  \left\langle\left( ( 2  /{\epsilon} )
\alpha' \left( { x_1}/{\epsilon} \right) (h D_{x_1})^2   + 
c ( x ) \alpha \left( { x_1}/{\epsilon} \right)  u  \right) , u \right\rangle 
+ \Im  \left\langle ( 2 h /{\epsilon}^2 )
 \alpha'' \left( { x_1}/{\epsilon} \right) h D_{x_1}   
u , u \right\rangle =  {\mathcal O} ( \epsilon^{\frac 12 } ) \,. 
\end{equation}
Integration by parts gives
\[ 
 \Im \left\langle ( 2 h /{\epsilon}^2 )
 \alpha'' \left( { x_1}/{\epsilon} \right) h D_{x_1}   
u , u \right\rangle = 
\Re \left\langle ( 2 h^2 /{\epsilon}^3 )
 \alpha''' \left( { x_1}/{\epsilon} \right) 
u , u \right\rangle
\,. \]
Using the  the fact that
$ P u = {\mathcal O}_{L^2} ( h ) $
we obtain 
\[ ( h D_{x_1})^2 u  = c ( x ) x_1 u - \lambda^w ( x , h D) u + {\mathcal 
O}_{L^2}  ( h) \,.\]
Hence 
from \eqref{eq:propa},
the nonnegativity of $\lambda ( x , \xi') $, we then see that the first 
term in \eqref{eq:form} satisfies
\begin{equation}
\label{eq:pos} 
\begin{split}
& \left\langle\left( ( 2  /{\epsilon} )
\alpha' \left( { x_1}/{\epsilon} \right) (h D_{x_1})^2   + 
c( x) \alpha \left( { x_1}/{\epsilon} \right)  u  \right) , u \right\rangle 
\\ & \ \ \ \ \ \geq \frac{1}{ C } \| u \|_{ L^2 ( \Omega_\epsilon ) } +  
\frac{1}{ C} \langle ( 1 + ( x_1/\epsilon )^2)^{-3/4} u , u \rangle 
- {\mathcal O} ( \epsilon^{\frac12} ) \,. 
\end{split}
\end{equation}
We used here $ \epsilon > h^{\frac23} $ which gave $ {\mathcal O} ( h/ \epsilon) 
= {\mathcal O} ( h^{1/3} ) = {\mathcal O} ( \epsilon^{1/2} ) $.

To estimate the second
term in \eqref{eq:form} 
we note that \eqref{eq:propa} gives
 $ \alpha''' ( t) = 0 $ for
$ t \leq 0 $ and $ | \alpha''' ( t ) | \leq C ( 1 + t^2)^{-7/4} $
for $ t > 0 $.
Using the assumption on $ \epsilon $, $ \epsilon > M h^{2/3} $, 
and choosing $M $ sufficiently large
we obtain 
\begin{equation*}
| \left\langle  h^2 {\epsilon}^{-3} 
 \alpha''' \left( { x_1}/{\epsilon} \right)     
u , u \right\rangle | 
\leq M^{-3} \langle ( 1 + ( x/\epsilon)^2 )^{-7/4} u, u \rangle  
\leq \frac{1}{ C} \langle ( 1 + ( x_1/\epsilon )^2)^{-3/4} u , u \rangle 
\end{equation*}
This and \eqref{eq:pos} show the second term in \eqref{eq:form} can be absorbed into
the first one. Taking this into account in 
combining \eqref{eq:form} and \eqref{eq:pos} completes the proof.
\end{proof}

The next lemma gives $ L^p $ estimates in strips. 
It follows the idea of \cite{KoT} of using rescaled Strichartz
estimates.
\begin{lem}
\label{l:7.3}
Suppose that
\[ A_\epsilon \defeq  \{ x \in \RR^n \,, \ \  
| x_1 - \epsilon | < \epsilon/2 \} \,, \]
and that $ u $ satisfies the localization condition \eqref{eq:loc}, 
$  u  = {\mathcal O}_{L^2} ( 1) $, 
$ P u = {\mathcal O}_{L^2} ( h ) $. Then 
\begin{equation}
\label{eq:cr}
 \| u \|_{ L^p ( A_\epsilon ) } = {\mathcal O} ( h^{-\sigma(p)}
\epsilon^{\frac14 -\mu( p) } ) \,, \ \ 2 \leq p 
\leq \frac{2n}{n-2} \,,  \ \ 
\epsilon > h^{2/3} \,, 
\end{equation}
where 
\[ \sigma ( p ) =  \left\{ \begin{array}{ll} 
\frac{n-1}{2} - \frac{n}{p} , &  \frac{2(n+1)}{n-1} 
\leq p \leq \infty ,\\
\ & \ \\
\frac{n-1}2\left(\frac12-\frac1p\right) , & 2 \leq p \leq
 \frac{2(n+1)}{n-1},  \end{array} \right. 
\]
and 
\[  \mu ( p ) = n \left( \frac 12- \frac1p \right) - \frac {3 \, \sigma ( 
p ) } 2 \,. \]
\end{lem}
\begin{proof}
Let us divide the strips into boxes of size $\epsilon$:
\[ \begin{split} 
& A_\epsilon^k \defeq  \{ x \in \RR^n \,, \ \  
| x_1 - \epsilon | < \epsilon/2 \,, \ \ 
| x' - \epsilon k |_{\ell^\infty}  < \epsilon/2 \} \,, \\  
& \widetilde A_\epsilon^k \defeq  \{ x \in \RR^n \,, \ \  
| x_1 - \epsilon | < 3\epsilon/4 \,, \ \ 
| x' - \epsilon k |_{\ell^\infty} 
 < 3\epsilon/4 \} \,, \ \ k \in \ZZ^{n-1} \,. 
\end{split} \]
We will prove that 
\begin{gather}
\label{eq:l73}
\begin{gathered}
\| u \|_{ L^p ( A_\epsilon^k ) } \leq 
C h^{-\sigma(p)}
\epsilon^{ -\mu( p) } ( \| u \|_{ L^2 ( \widetilde A_\epsilon^k) }
+ \epsilon^{\frac 12}  \| P u \|_{ L^2 ( \widetilde A_\epsilon^k) } ) 
  \,, \\ 2 \leq p \leq \frac {2n} {n-2} \,,  \ \ 
\epsilon > h^{2/3} \, . 
\end{gathered}
\end{gather}
As $ p \geq 2 $, and $ P u = {\mathcal O}_{L^2} ( h ) $,
\[ \begin{split} 
\| u \|_{  L^p ( A_\epsilon ) } & \leq 
C h^{-\sigma(p)}
\epsilon^{ -\mu( p) } \left( 
\sum_{ k
 } \left( \| u \|_{ L^2 ( \widetilde A_\epsilon^k) } + 
 (\epsilon^{\frac12} / h ) 
 \| P u \|_{ L^2 ( \widetilde A_\epsilon^k) }  \right)^p 
\right)^{\frac 1 p} \\
&  \leq C' h^{-\sigma(p)}
\epsilon^{ -\mu( p) } \left( 
\sum_{ k 
  }\left( \| u \|_{ L^2 ( \widetilde A_\epsilon^k) }^2 + 
( \epsilon /h^2) \| P u \|_{ L^2 ( \widetilde A_\epsilon^k) }^2  \right)
\right)^{\frac 1 2} \\
& \leq C'' h^{-\sigma(p)}
\epsilon^{ -\mu( p) }(  \| u \|_{ L^2 ( \Omega_{2 \epsilon } )} 
+  \epsilon^{\frac12}  )  \,,
\end{split} \]
from which the lemma follows by applying Lemma \ref{l:73}.

To prove \eqref{eq:l73} we rescale variables in $ \widetilde A_\epsilon^k$:
\[ \tilde x_1 \defeq ( x_1 - \epsilon )/ \epsilon \,, \ \ 
\tilde x' \defeq ( x' - \epsilon k ) / \epsilon \,,\]
and define
\[
 \tilde u ( \tilde x ) \defeq \epsilon^{\frac n 2} u ( \epsilon + \epsilon 
\tilde x_1 , \epsilon k + \epsilon x' ) \,, \ \
\widetilde P \tilde u \defeq \frac{1}{ \epsilon } \widetilde { P u } \,.\]
The operator $ \widetilde P $ is a semiclassical operator with 
a new parameter:
\[ \widetilde P = ( \tilde h D_{ \tilde x_1} )^2 + 
\tilde \lambda^w ( \tilde x , \tilde h D_{ \tilde x'} , \tilde h 
) - \tilde c ( 
\tilde x , \tilde h ) \tilde x_1 \,, \ \  \tilde h = h/ \epsilon^{3/2} \,. \]
Let 
\[ \begin{split}
&  
\widetilde A \defeq \{ \tilde x \; : \; | \tilde x_1 
- 1 | < 3/4 \,, \ 
| \tilde x' |_{ \ell^\infty } < 3/4  \} \,, \ \ 
A \defeq \{ \tilde x \; : \; | \tilde x_1 - 1| < 1/2 \,, \ \
 | \tilde x' |_{ \ell^\infty } \leq 1/2 \} \,. 
\end{split} \]
\setcounter{equation}{13}
By rescaling the desired estimate \eqref{eq:l73} is equivalent to 
\begin{equation}
\label{eq:resc}
 \| \tilde u \|_{ L^p ( A ) } \leq C \tilde h ^{- \sigma ( p ) } 
( \| \tilde u \|_{ L^2 ( \widetilde  A) } + 
 \tilde h ^{-1} \| \widetilde P \tilde u \|_{ L^2 ( \widetilde  A) } ) 
\,, \ \  2 \leq p \leq   \frac {2n} {n-2} \,. 
\end{equation}
In fact, $ \mu ( p ) = n ( 1/2 - 1/p) - 3 \sigma ( p ) /2 $, where
$ n ( 1/2 - 1/p ) $ comes from converting $ \tilde x $ integration to 
$ x $ integration. 

Using the elliptic estimate in Lemma \ref{l:ell}
(with $ h $ replaced by $ \tilde h $) we only need to prove 
\eqref{eq:resc} with
$ \tilde u $ supported in $ \widetilde A $: if 
$ \psi \in \CIc ( \tilde A ) $,
$ \psi = 1 $ in $ A $, then 
\[ \begin{split}  \| \tilde u \|_{ L^p ( A ) } & = 
 \| \psi \tilde u \|_{ L^p ( A ) } \leq C \tilde h ^{- \sigma ( p ) } 
( \| \psi\tilde u \|_{ L^2 ( \widetilde  A) } + 
 \tilde h ^{-1} \| \widetilde P\psi  \tilde u \|_{ L^2 ( \widetilde  A) } ) 
\\
&  \leq C' \tilde h ^{- \sigma ( p ) } 
( \| \tilde u \|_{ L^2 ( \widetilde  A) } + 
 \tilde h ^{-1} \| \widetilde P \tilde u \|_{ L^2 ( \widetilde  A) } ) 
\end{split} \]

We now observe that for $ \tilde x \in \tilde A $, $ \widetilde P $ 
satisfies the assumptions of Theorem \ref{th:seso}, with the new
semiclassical parameter $ \tilde h $. However, $ \tilde u $ does
not satisfy the localization condition \eqref{eq:loc} (again with 
$ \tilde h $). To remedy this, let $ \chi \in \CIc ( T^*\RR^n ) $ be
equal to one near 
\[ \bigcup_{ 0 \leq 
\tilde h \leq \tilde h_0 } \{ ( \tilde x , \tilde \xi) \; : \; 
\tilde x \in \widetilde A \,, \ \ 
\tilde p ( \tilde x , \tilde \xi, \tilde h ) = 0 \} \,, \ \ 
\widetilde P = \tilde p^w ( \tilde x , \tilde h D_{\tilde x } , 
\tilde h ) \,,\]
where we note that the definition of $ \widetilde P $ guarantees
the compactness of the union. Then $ \tilde u_1 \defeq 
\chi^w ( \tilde x , \tilde h D_{ \tilde x } ) \tilde u $, 
satisfies \eqref{eq:loc} (with $ h $ replaced by $ \tilde h $).
We can apply Theorem \ref{th:seso}, or rather its interpolated version,
shown in Fig.\ref{fig}, to see that
\[ \begin{split} \| \tilde u_1 \|_{L^p( A ) } & \leq 
C \tilde h^{-\sigma ( p ) } ( \| \tilde u_1 \|_{ L^2 (  \widetilde A) }
+ \tilde h^{-1} \| \tilde P \tilde u_1 \|_{ L^2({ \widetilde  A }) } ) 
\\ & \leq 
C \tilde h^{-\sigma ( p ) } ( \| \tilde u \|_{ L^2 ( \widetilde A ) }
+ \tilde h^{-1} \| \tilde P \tilde u \|_{ L^2 ({ \widetilde A }) } ) \,.
\end{split} \]
Here we also used 
Lemma \ref{l:ell} (with $ h $ replaced by $ \tilde h $) to estimate
the commutator terms arising in replacing 
$ \tilde u_1 $ with $ \tilde u $ on the right hand side.

We need to estimate $ \| \tilde u_2 \|_{ L^p (  A) } $, where
$ \tilde u_2 \defeq ( 1 - \chi )^w( \tilde x , \tilde h D_{\tilde x} ) 
\tilde u $.
For that we note that on the support of $ 1 - \chi $, $ \tilde p \geq 
\langle \tilde \xi \rangle ^2/ C  $, that is we have strong ellipticity.
We can apply Lemma \ref{l:1} to obtain
\[  \sum_{ | \alpha | \leq 2 } \| ( \tilde h D_{\tilde x } )^\alpha 
\tilde u_2 \|_2  \leq C \| \widetilde P \tilde u_2 \|_{ L^2 
( \widetilde A ) } + {\mathcal O} ( \tilde h^\infty ) \| \tilde u 
\|_{ L^2 ( \widetilde  A ) } \,. \]
Lemma \ref{l:Sob} now shows that 
\[ \| \tilde u_2 \|_{p } \leq C \tilde h^{1-n(1/2 - 1/p)} 
(  \| \tilde u \|_{ L^2 ( \widetilde  A ) } + 
(1/\tilde h)  \| \widetilde P \tilde u \|_{ L^2 
( \widetilde A ) } )  \,,  \ \ 
\frac12 - \frac 2 n \leq \frac 1 p \leq \frac 12 \,, \ \ p < \infty \,.\]
We note that 
except for $ n = 2 $, 
the condition on $ p $ is the same as the condition in 
\eqref{eq:resc} and that $  n ( 1/2 - 1/p ) -1 \leq \sigma ( p ) $. 
When $ n = 2 $ we have to consider the case of $ p = \infty $, and 
the same estimate follows from Lemma \ref{l:Ber} applied with $ s = 2 $.

Thus for all 
$ n \geq 2 $ 
we obtained a stronger version of \eqref{eq:resc} with 
$ \tilde u $ replaced by  
$ \tilde u_2 $ on the right hand side 
(we could not directly invoke Theorem \ref{th:ell} 
since we do not have localization condition for $ \tilde u_2 $).

Writing $ \tilde u  = \tilde u_1 + \tilde u_2 $ and combining the
two estimates give \eqref{eq:resc} proving the lemma.
\end{proof}

\noindent
{\em Proof of Theorem \ref{th:newg}:} 
Using Lemma \ref{l:7.3} 
we obtain the estimate in $ \RR^n \setminus \Omega_{ h^{2/3} } $
by using a dyadic decomposition with $ \epsilon = 2^k h^{2/3} $.
We check that in \eqref{eq:cr} we have 
\[\begin{split}
&  \mu \left( \frac { 2 ( n + 3 ) }{ n + 1 } \right) = \frac 1 4 \,, \ \
- \mu ( p ) + \frac 14 > 0 \,, \ \ 2 \leq p < \frac { 2 ( n + 3 ) }
{ n + 1 }  \,, \\
&  \mu \left( \frac { 2  n   }{ n - 1 } \right) = \frac 1 4 \,, \ \
- \mu ( p ) + \frac 14 < 0 \,, \ \   \frac { 2 ( n + 3 ) }
{ n + 1 } < p  < \frac{2n }{ n-2} \,.
 \end{split} \]
Hence, with $ K ( h ) = {\mathcal O} ( \log ( 1/h ) ) $ 
given by  $ 2^{K(h) } = h^{2/3} $,
\[ \begin{split}
 \| u \|^p_{L^p (  \RR^n \setminus \Omega_{ h^{2/3} } ) }
& \leq C \sum_{ k =0 }^{K ( h ) } 
 h^{-p( \sigma(p) + 2/3(\mu( p ) - 1/4) )    } 
2^{pk(1/4-\mu( p )) }   \\
& = 
 \left\{ \begin{array}{ll}
{\mathcal O} ( h^{- p ( 2/3) ( n ( 1/2 - 1/p) - 1/4) } ) =
{\mathcal O} 
( h^{ p \left( \frac 1 6  - \frac {2 n } 3 ( 1/2 - 1/p )  \right) } ) & 
\frac{2(n+3)}{n+1} < p < \frac{2n}{n-2} \,, \\
\ & \ \\
{\mathcal O} ( h^{- p \sigma(p)  } K ( h ) ) = 
{\mathcal O} ( h^{- p (n-1)/(2(n + 3))  }  \log ( 1/ h ) ) & 
p = \frac{2(n+3)}{n+1} \,, \\
\ & \ \\
{\mathcal O} ( h^{- p \sigma(p)  } ) =
{\mathcal O} ( h^{- p ( ( n-1 ) ( 1/2 - 1/p ) ) /2  } )
 & 
2 \leq p \leq  \frac{2(n+3)}{n+1}  \,,
\end{array} \right.
\end{split} \]
which is the desired estimate \eqref{eq:thnew} for $ 
2 \leq p < 2n / ( n - 2 ) $. When $ n > 2 $ the estimate
for $ p = 2n / ( n -2 ) $ follows from Theorem \ref{th:nonpr}.
When $ n = 2 $ then the $ L^\infty $ estimates follows 
from the estimate in strips given in Lemma \ref{l:7.3}.

To complete the proof we estimate the norm of the truncated 
function $ u_\epsilon $, 
which appeared already in the proof of Lemma \ref{l:7.1}:
\[ u_\epsilon = \chi ( x_1/\epsilon ) u \,, \ \
 \chi \in \CI ( ( - \infty , 2) , [ 0 , 1] ) \,, \ \
\chi ( t ) =1 \,, \  t \leq 1 \,, \ \ 
 \epsilon = h^{2/3} \,. \]
Lemma \ref{l:7.3} then shows that
\[ \sum_{ | \alpha | \leq 1 } \| ( h^{2/3} D )^{\alpha } u_\epsilon \|_2 
= {\mathcal O} ( h^{\frac 16 } ) \,.\]
Applying Lemma \ref{l:Sob} with $ h $ replaced by $ h^{2/3} $ we see that
\[ \| u \|_{L^p ( \Omega_{h^{2/3} } ) } \leq
 \| u_\epsilon \|_p \leq  
C h^{(2/3)n(1/p-1/2) + 1/6}  \,, \ \  2 \leq p < \frac { 2n } { n - 2} 
\,. \]
This completes the proof for $ p > 2n/(n-2) $ as the last 
 estimate is the same as \eqref{eq:thnew} for 
$ 2 ( n + 3 ) / ( n + 1 ) < p < 2n / ( n - 2 ) $
and better for the remaining values of $ p$. For $ n > 2 $ the 
result at $ p = 2n/(n-2) $ again follows from Theorem \ref{th:nonpr}.

For $ n = 2 $ we recall from the proof of Lemma \ref{l:7.1} 
that for $ \epsilon = h^{2/3} $, 
\[ \| P u_\epsilon \|_2 = {\mathcal O} (1 ) \| P u \|_2 + 
{\mathcal O} ( h^{\frac13} ) \| h D_{x_1} u \|_{\Omega_{2\epsilon} }
+ {\mathcal O} ( h^{\frac23} ) \| u\|_{ \Omega_{2\epsilon } } \,, \]
and hence by Lemma \ref{l:7.3}, $ \| P u_\epsilon \|_2 = {\mathcal O}
( h^{5/6} ) $. We now recall \eqref{eq:formp} and write
\[ \begin{split} \| P u_\epsilon \|_2^2 & = 
\|( ( h D_{x_1} )^2 + a ( x ) ( hD_{x_2} )^2 )u_\epsilon \|^2 _2 
+ \| c ( x ) x_1 u_\epsilon \|^2_2 \\ & \ \ \ \ \ \ \ \ \ \ -  
2 \Re \langle c ( x ) x_1 u_\epsilon , 
( ( h D_{x_1} )^2 + a ( x ) ( hD_{x_2} )^2 )u_\epsilon  \rangle \,.
\end{split}
\]
The last term is equal to 
\[ \begin{split}  & - 2 \Re \langle c ( x) x_1 h D_{x_1} u_\epsilon , 
h D_{x_1} u_\epsilon \rangle - 
 2 \Re \langle c ( x) a ( x ) x_1 h D_{x_2} u_\epsilon , 
h D_{x_2} u_\epsilon \rangle \\ 
& \ \ \ \ \ \ \ \ + {\mathcal O}( h ) \| u_\epsilon \|_2 ( 
\| h D_{x_1} u_\epsilon \|_2 + 
\| h D_{x_2} u_\epsilon \|_2 ) \\
&  \ \ \ \ \ \geq - {\mathcal O } ( h^{5/3} ) \,, \end{split} \]
where we used $ x_1 \leq 2h^{2/3} $  and Lemma \ref{l:7.3}.
Hence
\[  \|( ( h D_{x_1} )^2 + a( x) ( hD_{x_2} )^2 )u_\epsilon \|_2 =
{\mathcal O} ( h^{\frac56} ) \,,  \]
and we obtain
\[  \|( ( h^{2/3} D_{x_1} )^2 + a ( x)( h^{2/3} 
D_{x_2} ) )u_\epsilon \|_2 =
{\mathcal O} ( h^{\frac16} ) \,, \]
Using Lemma \ref{l:ell} we consequently have
\[ \sum_{ | \alpha | \leq 2 } \| ( h^{2/3} D )^{\alpha } u_\epsilon \|_2 
= {\mathcal O} ( h^{\frac 16 } ) \,.\]
Finally, Lemma \ref{l:Ber} shows that
\[ \| u \|_{ \infty } = {\mathcal O} ( h^{\frac16 - \frac23 } ) 
=  {\mathcal O} ( h^{-\frac12} ) \,, \]
completing the proof for $ n =2 $, $ p = \infty $.

\stopthm

\end{document}